\documentclass[11pt,a4paper]{amsart}

\usepackage[latin1]{inputenc}
\usepackage[english]{babel}
\usepackage{amsmath}

\usepackage{amssymb, mathabx}
\usepackage[numbers]{natbib}
\usepackage{graphicx}

\usepackage{braket}
\usepackage[colorlinks,hyperindex,bookmarksopen,linkcolor=red,citecolor=blue,urlcolor=blue]{hyperref}

\usepackage{mathrsfs}

\usepackage{braket}

\bibpunct{[}{]}{;}{n}{,}{,}
\usepackage[hmargin=3cm,vmargin={3.5cm,4cm}]{geometry}

\theoremstyle{theorem}

\theoremstyle{definition}                                 

\theoremstyle{definition}                           

\theoremstyle{remark}                             

\usepackage{color}

\usepackage{mathtools,slashed}

\newcommand{\be}{\begin{eqnarray}}
\newcommand{\ee}{\end{eqnarray}}

\numberwithin{equation}{section}

\allowdisplaybreaks

\begin{document}

\title{Modeling sorption of emerging contaminants in biofilms}

\author{Luigi Frunzo}
\address{University of Naples "Federico II", Department of Mathematics and
					 Applications, via Claudio 21, 80125, Naples, Italy\
          \texttt{luigi.frunzo@unina.it} }


\date  {\today}

\begin{abstract}

	A mathematical model for emerging contaminants sorption in multispecies biofilms, based on
	a continuum approach and mass conservation principles is presented. 
	Diffusion of contaminants within the biofilm is described using a diffusion-reaction equation. 
  Binding sites formation and occupation are modeled by two systems of hyperbolic partial differential
  equations are mutually connected through the two growth rate terms. The model is completed with a 
	system of hyperbolic equations governing the microbial species growth within the biofilm; a system 
	of parabolic equations for substrates diffusion and reaction and a nonlinear ordinary differential 
	equation describing the free boundary evolution.
  Two real special cases are modelled. The first one describes the dynamics of a free sorbent
	component diffusing and reacting in a multispecies biofilm. In the second illustrative case, the
	fate of two different contaminants has been modelled. 
		%
\end{abstract}

\maketitle

\section{Introduction} \label{n1}
 \setcounter{equation}{0} \setcounter{figure}{0}

 \textit{Biofilms} can be defined as a form of microbial ecosystem constituted by colonies of microorganisms, embedded in a
   primarily polysaccharides matrix and attached to a solid surface or in suspensions, e.g. flocs or granules.
 The biofilm matrix is composed by extracellular material, mostly self-produced by the organisms.
 This material consists of a conglomeration of different types of bio-polymers, known as extracellular 
 polymeric substances (EPS). 
 The EPS, and thus, the composition of the biofilm itself, can vary greatly among biofilms,                   
 depending on the microbial species constituting the system, the temperature and the availability of nutrients.
 Overall, the biofilm matrix is characterized by the presence of proteins, lipids, and the products of
 bacterial lisys and decay such as nucleic acids, or external DNA \cite{flemming2010biofilm,schleheck2009pseudomonas}. 
 
 The heterogeneity of the biofilm matrix provides to the biofilms several properties. These properties can be classified as
 mechanical: adhesion to a surface, aggregation of bacteria cells, protective barrier;
 biological: digestion of exogenous macromolecules, nutrient storage; and physico-chemical: 
 sorption of organic compounds, and inorganic ions.
 The latter allow to use the biofilm matrix as a biosorbing agent by accumulating xenobiotics
 or toxic metal ions from the environment. This process known as biosorption can be defined as the complex combination of processes
 aimed at the entrapment of a substance onto the surface of a living/dead organism or on EPS \cite{gadd2009biosorption,da2016copper}.
 
 In the last years there has been a growing interest in a new class of contaminants, the so called \textit{emerging contaminants}.
 These compounds that have only recently been categorised as contaminants, include a wide array of substances such
 as xenobiotic ( pharmaceuticals and personal care products), pesticides, veterinary products,
 industrial compounds/by-products, food additives, engineered nano-materials, and heavy metals ions \cite{lapworth2012emerging}.
 During the last years, extensive experimental studies have been carried out on the
 immobilization of some categories of \textit{emerging contaminants} by biosorption \cite{gadd2009biosorption,d2015mathematical}.
 Many biological materials are suitable for maintaining biosorption due to the high efficiency, cost
 effectiveness and particular affinity with these pollutants \cite{gadd2009biosorption}. 
 The potential of microbial biomass as biosorbents has been largely studied and reviewed \cite{van2003metal}. 
 Moreover, since most microorganisms live in form of biofilms, the different nature of the cell
 agglomerate and the heterogeneous composition of biofilm matrix further contribute to biosorption \cite{flemming2010biofilm}.

 Despite the growing interest of the scientific community, the application of biosorption at the
 industrial scale has not been yet exploited, mainly due to the complexity of the mechanisms
 involved in this technique. 
 Therefore, a mathematical model appears as a support tool to gain essential information for
 the identification of the key factors affecting biosorption efficiency and stability \cite{d2015mathematical}.
 
 Mathematical researches on biofilms show increasing trend in the last years, due
 also to their importance in engineering, biological and industrial applications \cite{de2009failure,de2010senescence,rahman2015mixed,AG-FCAA-2017}. 
 In spite of this extensive modeling activity, little attention has been directed towards
 mathematical modeling of sorption phenomena in multispecies biofilms. The first attempt, in the
 framework of continuous models, was recently presented in \cite{d2015mathematical}, where the authors proposed a
 mathematical model for EPS metal biosorption that describes biofilm growth dynamics, including
 the spatial distribution of microbial species, substrate concentrations and EPS formation.
 The authors modeled the diffusion
 of the free metal ions within the biofilm and their further sorption on the EPS biofilm component by using
 a diffusion-reaction equation based on the coupled diffusion-adsorption approach for thin film
 firstly introduced in \cite{bartlett1996diffusion}. 
 
\begin{figure}
\centering
\includegraphics[width=1\textwidth]{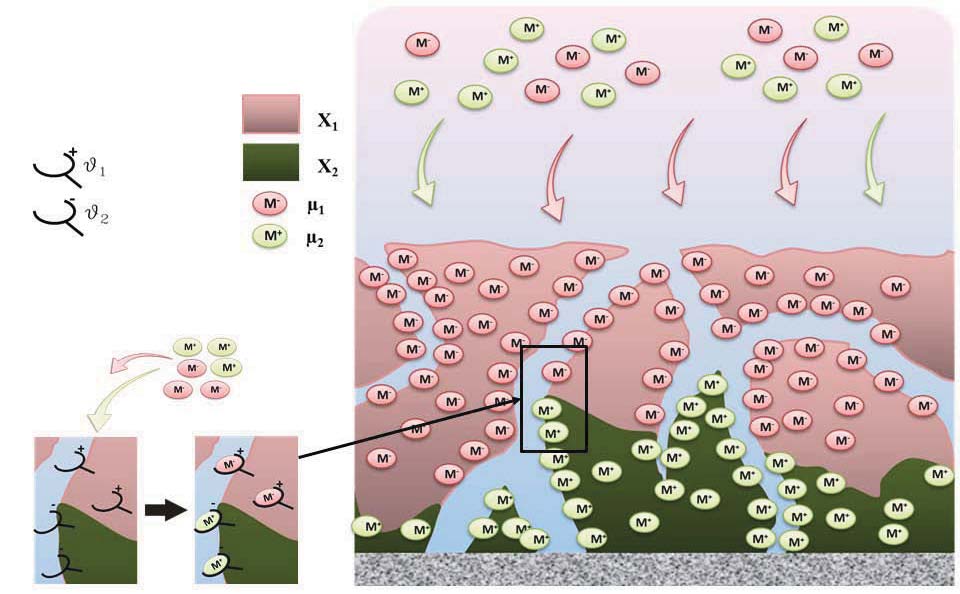}
\caption{Schematic representation of biosorption process.}
         \label{f1.2}     
\end{figure}

 In this proposed work, a general model for the sorption of \textit{emerging contaminants} on the several
 components, EPS, active biomass
 and inert, constituting the biofilm matrix is introduced. In particular, the hyperbolic partial differential equations
 describing the dynamics of the free binding sites and the occupied binding sites, for each biofilm components and
 the parabolic partial differential equations, describing the diffusion-reaction of the contaminants
 within the biofilm have been derived.
 These equations have been coupled with the equations describing biofilm growth and substrate uptake in order to define
 the complete problem. The four systems of partial differential equations result strictly connected through the reaction
 terms, defined in this work.
 It is interesting to note that the coupled diffusion-adsorption model introduced in \cite{bartlett1996diffusion}
 represents a special case of the mathematical model introduced in this work [\textbf{Remark1}].
 The approach is based on a continuum model in one space dimension and then generalized to three
 dimension with the intention of predicting biofilm growth, spatial distribution of microbial 
 species, substrate trends, attachment and detachment, and in particular the formation of new
 free binding sites and occupied binding sites, the diffusion of sorbent contaminants, and the
 spatial distribution of sorbed contaminant in the biofilm. 
 The empirical observation shows that the biofilm has a complex
 and heterogeneous structure and each component is characterized by the
 presence of selective binding sites for a specific sorbent contaminant.
 The model takes into account the sorption of a specific contaminant $\mu_i$ on a specific binding
 site $\vartheta_i$ characteristic of the related biofilm component $X_i$, Figure \ref{f1.2}. 
 The formation of new (free) binding sites has been related to the formation of new biofilm matter. 
 The occupation of the free binding sites, and thus the formation
 of occupied sites has been incorporated in the model by considering a dependance of the sorption rate
 on the concentration of free sorbent contaminants and free binding sites.
 The diffusion and reaction of free sorbent contaminants has been described by using a diffusion-reaction
 equation, supposing a Fick's law diffusion and connecting the reaction of contaminants with
 the occupation of binding sites.  
 Notably, the formation of new binding sites has been modeled by a hyperbolic partial differential
 equations (PDEs) system characterized by the presence of two reaction terms. The first one, related to the bio-conversion
 of the substrates, models the formation of the new binding sites. The second one represents the occupation
 rate of the free binding sites.    
 The occupation of the free sites is modeled by a system of hyperbolic PDEs as well. This system
 is connected to the previous hyperbolic equations through the reaction term which is the opposite
 of the occupation rate previously defined.   
 The diffusion and reaction of the free sorbent contaminant is described by a system of parabolic PDEs where
 the reaction terms model the sorption rate of the free sorbent contaminants on the free sites. 
 The sorption rate and the occupation rate are connected through a yield and the density of occupation binding sites.  
 Notably, the equations introduced in this work follow from mass conservation principles.

 Finally special models have been defined in order to model real cases. In particular, two cases have been considered.
 The first one (Case I) describes the dynamics of
 a free sorbent component diffusing in a multispecies biofilm. This application is organized in two parts,
 case Ia and case Ib. Case Ia is focused on the effect of the binding sites density
 on the sorption phenomenon. In application Ib the effect of different adsorption constants has been evaluated. 
 In the second special case, case II, the fate of two different contaminants has been evaluated. In particular,
 the adsorption of two different sorbent compounds on two biofilm components has been evaluated. 
 Numerical simulations confirm the capability of the model to predict sorbed contaminants distribution into the biofilm, 
 free and occupied binding sites formation, percentage of occupied sites, biomass distribution, substrate concentration
 profile over biofilm depth and biofilm thickness.
 
 The paper is organized as follows. In Section 2, in the framework of continuum mechanics, the equation governing
 the described phenomena are derived by following conservation mass principles. Section 3  introduces the complete
 model and specifies the reaction terms for each equation.  Section 4 describes the two experimental cases to which
 the new model is applied and presents the numerical results. Finally, in Section 5 we discuss the conclusions.


 \section{Mathematical Model}   \label{n2}

 \subsection{Equation for biofilms and substrates} \label{n2.1}
 
 In the framework of continuum approach, the evolution of biofilm is modeled through
 the concentration of the biofilm components, $X_i(z,t), i=1,...,n$, 
 and the concentration of the substrates, $S_j(z,t), j=1,...,m$.
 By considering multispecies biofilm growth in one space dimension and denoting by $z$ the biofilm growth direction,
 assumed perpendicular to the substratum located at $z=0$, the dynamics of the system are governed by the
 following equation:

 \begin{equation}                                        \label{2.1}
 \frac{\partial X_i}{\partial t}
 +\frac{\partial}{\partial z}(u X_i)
 =\rho_i r_{M,i}, \ \ i=1,...,n,\ 0\leq z\leq L(t),\ t>0,
 \end{equation}
 
 where  $\rho_i$ denotes constant density, $L(t)$ denotes biofilm thickness, free boundary;
 $r_{M,i}$ is the specific growth/formation rate; $X_i=\rho_if_i$ where $f_i$ is the volume fraction
 of biofilm particulate components $i$.

 $u(z,t)$ is the velocity of the microbial mass, governed by the following equation

 \begin{equation}                                        \label{2.2}
 \frac{\partial u}{\partial z}
 =\sum_{i=1}^{n} r_{M,i},\ 0< z\leq L(t),\ t\geq0,
 \end{equation}

 The differential equation for $L(t)$ is obtained by the global mass balance and gives:

 \begin{equation}                                        \label{2.3}
   \dot L(t)=u(L(t),t)+\sigma_a^{}(t)-\sigma_d^{}(L(t)),\ t>0,
 \end{equation}

 where $\sigma_a^{}(t)$  is the attachment biomass flux from bulk liquid to biofilm and 
 $\sigma_d^{}(L(t))$ denotes the detachment biomass flux from biofilm to bulk
 liquid.
 The hyperbolic partial differential equations (\ref{2.1}) describe the evolution, in space and time, of
 microbial species in the biofilm as a convective flux regulated by the bio-conversion of organic substrates.
 The equation system (\ref{2.1}) derives from local mass balance. It was first presented in \cite{art:rif.8} 
 and introduced in the general form above in \cite{art:rif.16}. 
 It is important to note that equation (\ref{2.1}) is generally used to describe the dynamics of 
 all the particulate components, constituting the biofilm, including inert biomass and EPS. 

 Diffusion and bioreaction of substrates within biofilms is governed by the following semi-linear parabolic equations

 \begin{equation}                                        \label{2.4}
   \frac{\partial S_j}{\partial t}-\frac{\partial}{\partial z}\left(
   D_j\frac{\partial S_j}{\partial z}\right)=
   r_{S,j}, \ \ 0< z<L(t),\ t>0,\ j=1,...,m,
 \end{equation}

 where $r_{S,j}$ is the conversion rate of substrate
 $j$, and $D_{S,j}$ denotes the diffusivity coefficient of substrate $j$. 

Initial-boundary conditions for equations (\ref{2.1}) and (\ref{2.2}) are prescribed as
  
	\begin{equation}                                        \label{2.5}
	X_i(z,0)= X_{i0}(z), \ u(0,t)=0, \ i=1,...,n, \  0\leq z\leq L_0, \ t\geq 0
	\end{equation}
	
	\noindent
	where the functions $X_{i0}(z), i=1,...,n$, represent the initial concentrations 
	of microbial species and equation (\ref{2.5})$_2$ is a no flux condition between substratum and biofilm.
	
	Suitable initial-boundary conditions for equation (\ref{2.4}) are: 
 \begin{equation}                                        \label{2.6}
   S_j(z,0)=S_{j0}(z),\ 0\leq z\leq L_0,
   \  j=1,...,m.
 \end{equation}
 No substrate flux is assumed at the substratum $z=0$,
 \begin{equation}                                        \label{2.7}
   \frac{\partial S_j}{\partial z}(0,t)=0,
   \ t>0,\ j=1,...,m.
 \end{equation}
 On the free boundary $z=L(t)$, Dirichlet conditions
 \begin{equation}                                        \label{2.8}
    S_j(L(t),t)=S_{jL}(t),\ t>0,\ j=1,...,m,
 \end{equation}
 or Neumann conditions
 \begin{equation}                                        \label{2.9}
    \frac{\partial S_j}{\partial z}(L(t),t)=S_{jL}(t),\ t>0,\ j=1,...,m,
 \end{equation}
 or mixed conditions can be prescribed based on the specific characteristics of the system to be modeled. 

In this work the conversion rate of substrate $r_{S,j}$ and the specific growth/formation
rate of the biofilm components depend also on  the concentration of the
sorbent contaminants $\mu_i(z,t)$, $i=1,...,n$,
\[
r_{S,j}=r_{S,j}(z,t,{\bf X},{\bf S},{\mbox{\boldmath $\mu$}}), \ \  j=1,...,m,
r_{M,i}^{}=r_{M,i}^{}(z,t,{\bf X},{\bf S},{\mbox{\boldmath $\mu$}}), \ \ i=1,...,n,
\]
where, ${\bf X}=(X_1,...,X_n)$, \ \ ${\bf S}=(S_1,...,S_m)$, \ \ ${\mbox{\boldmath $\mu$}}=(\mu_1,...,\mu_n)$.
The key role of $\mu_i$ in the dynamics of the system is apparent as the growth of the biofilm and the uptake
of the substrates depend on $\mu_i$. As mentioned in section \ref{n1}, the dynamics of $\mu_i$ depend not only
on microbial components and substrates, but also on the binding sites evolution.
The equations for the sorbent contaminants $\mu_i$ and for the binding sites will be derived and introduced in the next sections.

\subsection{Equations for binding sites}\label{n2.2}

\noindent
 Consider a control volume $(z_2-z_1)A$, with $A$ representing a constant cross-sectional area, and denote with
 $w_i(z,t)=u(t,z)N_i\vartheta_i(t,z)$ the biomass flux. The mass balance for the binding sites $i$ ($i=1,...,n$), is

 \[
   A\frac{\partial}{\partial t}\int_{z_1}^{z_2}N_i \vartheta_i\ dz
    =A[w_i(z_1,t)-w_i(z_2,t)]+
   A\int_{z_1}^{z_2} N_i (r_{M,i} - r_{D,i})\ dz,
 \]
 \begin{equation}                                        \label{2.10}
 \int_{z_1}^{z_2} \frac{\partial \vartheta_i}{\partial t}\ dz=
 -\int_{z_1}^{z_2} \frac{\partial (u\vartheta_i)}{\partial z}\ dz+
  \int_{z_1}^{z_2} r_{M,i}\ dz - 
	\int_{z_1}^{z_2} r_{D,i}\ dz,
 \end{equation}

 \noindent
  where the product $N_i\vartheta_i$ is the free binding sites concentration, with $N_i$ that represent the free binding site density.
	 \noindent
 Differentiation of equation (\ref{2.10}) with respect to $z_2$ and setting
 $z_2=z$ leads to
 
\begin{equation}                                        \label{2.11}
 \frac{\partial \ \vartheta_i}{\partial t}
 +\frac{\partial}{\partial z}(u\vartheta_i)
 = r_{M,i}(z,t,{\bf X},{\bf S},{\mbox{\boldmath $\mu$}})- r_{D,i}(z,t,{\mbox{\boldmath $\mu$}},{\mbox{\boldmath $\vartheta$}},{\mbox{\boldmath $\bar{\vartheta}$}}). 
 \end{equation}

\noindent
Eq (\ref{2.11}) models the fate of the free binding sites into the biofilm.
By using similar reasoning it is possible to obtain the following equation

\begin{equation}                                        \label{2.12}
 \frac{\partial \ \bar{\vartheta_i}}{\partial t}
 +\frac{\partial}{\partial z}(u\bar{\vartheta_i})
 = r_{D,i}(z,t,{\mbox{\boldmath $\mu$}},{\mbox{\boldmath $\vartheta$}},{\mbox{\boldmath $\bar{\vartheta}$}}).
 \end{equation}

\noindent
that models the fate of the occupied binding sites fractions, $\bar{\vartheta_i}$ into biofilm.

The following initial conditions are considered for Eq. (\ref{2.11}) and Eq. (\ref{2.12})
  
	\begin{equation}                                        \label{2.13}
	\vartheta_i(z,0)= \vartheta_{i0}(z), \ \bar{\vartheta_i}(z,0)= \bar{\vartheta_{i0}}(z), \ u(0,t)=0, \ i=1,...,n, \  0\leq z\leq L_0, \ t\geq 0
	\end{equation}

	\noindent
	where the functions $\vartheta_{i0}(z)$ and $\bar{\vartheta_{i0}}(z), i=1,...,n$, represent the initial binding site
	concentrations and eq. (\ref{2.13})$_3$ is a no flux condition between substratum and biofilm.

\subsection{Equations for contaminants}\label{n2.4}

Consider a control volume $(z_2-z_1)A$, with $A$ representing a constant cross-sectional area, and denote with
 $w_i(z,t)$ the emerging contaminant flux. The mass balance for $\mu_i$, $i=1,...,n$, is

 \[
   A\frac{\partial}{\partial t}\int_{z_1}^{z_2} \mu_i\ dz
    =A[w_i(z_1,t)-w_i(z_2,t)]-
   A\int_{z_1}^{z_2} Y_{ADS}N_ir_{D,i}\,
 \]
 \begin{equation}                                        \label{2.14}
 \int_{z_1}^{z_2} \frac{\partial \mu_i}{\partial t}\ dz=
 -\int_{z_1}^{z_2} \frac{\partial w_i}{\partial z}\ dz-
   \int_{z_1}^{z_2} Y_{ADS}N_ir_{D,i}\ dz,
 \end{equation}

 \noindent
  where $r_{D,i}$ represents the adsorption rate of \textsl{emerging contaminants} $\mu_i$. 
 
 \noindent
 Differentiation of equation (\ref{2.14}) with respect to $z_2$ and setting
 $z_2=z$ leads to
 \begin{equation}                                        \label{2.15}
 \frac{\partial \mu_i}{\partial t}
 +\frac{\partial}{\partial z}(w_i)
 =Y_{ADS}r_{D,i}(z,t,{\mbox{\boldmath $\mu$}},{\mbox{\boldmath $\vartheta$}},{\mbox{\boldmath $\bar{\vartheta}$}}).
 \end{equation}

\noindent
  According to Fick's first law, the flux of emerging contaminant $i$ within the
	biofilm is proportional to the diffusivity $D_{C,i}$ and may be expressed as

 \begin{equation}                                        \label{2.16}
    w_i= - D_{C,i} \frac{\partial \mu_i}{\partial z}.
 \end{equation}

 Substituting eq. (\ref{2.16}) into equation (\ref{2.15}) leads to

 \begin{equation}                                       \label{2.17}   
\frac{\partial \mu_i}{\partial t}-\frac{\partial}{\partial z}\left(
   D_{C,i}\frac{\partial \mu_i}{\partial z}\right)=
   - Y_{ADS}N_ir_{D,i}(z,t,{\mbox{\boldmath $\mu$}},{\mbox{\boldmath $\vartheta$}},{\mbox{\boldmath $\bar{\vartheta}$}}), \ \ 
	i=1,...,n,\ 0< z<L(t),\ t>0.
 \end{equation}

 \noindent
 where:
 
 \noindent
  
 ${\mbox{\boldmath $\mu$}}=(\mu_1,...,\mu_n)$;
 
 \noindent
 $Y_{ADS}$ is the yield of the contaminant $\mu_i$ and Occupied binding sites $\bar{\vartheta}_i$;
 
 \noindent
 $D_{C,i}$ denotes the diffusivity coefficient of contaminant $i$;

 Equations (\ref{2.17}) govern the diffusion and reaction through sorption of a special contaminant $\mu_i$ within biofilms. 
 The sorption mechanisms on the biofilm components (e.g active biomass, inert residual, and EPS) may differ qualitatively
 and quantitatively. This is taken into account by considering different adsorption rates and mechanisms modeled
 by $r_{D,i}(z,t,{\mbox{\boldmath $\mu$}},{\mbox{\boldmath $\vartheta$}},{\mbox{\boldmath $\bar{\vartheta}$}}))$. 
 Initial and boundary conditions for equations (\ref{2.17}) have to be prescribed.
 
 Typical conditions are

 \begin{equation}                                        \label{2.18}
   \mu_i(z,0)=\mu_{i0}(z),\ 0\leq z\leq L_0,
   \  i=1,...,n,
 \end{equation}

 \begin{equation}                                        \label{2.19}
   \frac{\partial \mu_i}{\partial z}(0,t)=0,
   \ t>0,\ i=1,...,n,
 \end{equation}
 
 \begin{equation}                                        \label{2.20}
    \mu_i(L(t),t)=\mu_{i L}(t),\ t>0,\ i=1,...,n,
 \end{equation}

 which reproduce the specific case of a contaminant $\mu_i$ present in the bulk liquid 
 at a given concentration $\mu_{i L}(t)$ and no substrate flux at the 
 substratum $z=0$.
 Different initial and boundary conditions may be also prescribed depending on the  
 specific problem discussed.

 \subsection{3D Model}\label{n2.5}

 The 1D model presented in the previous sections can be generalized to
 3D by starting from the model described in \cite{art:rif.10}. Denote
 by $B_t$ the 3D region occupied by the biofilm and let
 ${\bf x}=(x_1^{},x_2^{},x_3^{})$ be a generic point. Then,
  $X_i=X_i({\bf x},t)$,
 $f_i=f_i({\bf x},t)$,
 $S_j=S_j({\bf x},t)$ ,
 ${\bf u}={\bf u}({\bf x},t)$,
 $\vartheta_i=\vartheta_i({\bf x},t)$,
 $\bar{\vartheta}_i=\bar{\vartheta}_i({\bf x},t)$,
 $mu_i=\mu_i({\bf x},t)$,

 If ${\bf u}=-\nabla p$, where $p$ denotes the pressure within the biofilm,
 the equations governing biofilm and substrate evolution are
 written as

 \begin{equation}                                        \label{2.21}
 \frac{\partial X_i}{\partial t}
 -\nabla\cdot(X_i\nabla p)
 =\rho_i r_{M,i}^{}(z,t,{\bf X},{\bf S},{\mbox{\boldmath $\mu$}}),
 \ {\bf x}\in B_t,
 \end{equation}

\begin{equation}                                        \label{2.22}
 \frac{\partial \vartheta_i}{\partial t}
 -\nabla\cdot(\vartheta_i\nabla p)
 =r_{M,i}({\bf x},t,{\bf X},{\bf S},{\mbox{\boldmath $\mu$}})- r_{D,i}({\bf x},t,{\mbox{\boldmath $\mu$}},{\mbox{\boldmath $\vartheta$}},{\mbox{\boldmath $\bar{\vartheta}$}}),
 \ {\bf x}\in B_t,
 \end{equation}

\begin{equation}                                        \label{2.23}
 \frac{\partial \bar{\vartheta}_i}{\partial t}
 -\nabla\cdot(\bar{\vartheta}_i\nabla p)
 =r_{D,i}({\bf x},t,{\mbox{\boldmath $\mu$}},{\mbox{\boldmath $\vartheta$}},{\mbox{\boldmath $\bar{\vartheta}$}}),
 \ {\bf x}\in B_t,
 \end{equation}

 \begin{equation}                                        \label{2.24}
 \nabla^2p
 =-\sum_{i=1}^{n}r_{M,i},\ {\bf x}\in B_t,
 \end{equation}

 \begin{equation}                                        \label{2.25}
   \frac{\partial \mu_i}{\partial t}-D_{M,i}\nabla^2 \mu_i
 =r_{\psi,i}({\bf x},t,{\mbox{\boldmath $\mu$}},{\mbox{\boldmath $\vartheta$}},{\mbox{\boldmath $\bar{\vartheta}$}}),
   \ {\bf x}\in B_t,
 \end{equation}

 \begin{equation}                                        \label{2.26}
   \frac{\partial S_j}{\partial t}-
   D_j\nabla^2 S_j=
   r_{S,j}({\bf x},t,{\bf X},{\bf S},{\mbox{\boldmath $\mu$}}),\ {\bf x}\in B_t.
 \end{equation}


 \section{Complete model} \label{n3}
 \setcounter{equation}{0} 

The free boundary value problem is now completely described by the following set of
 differential equations: the hyperbolic eqs. (\ref{2.1}) for the biofilm
 volume fractions and for binding sites evolutions (\ref{2.6})-(\ref{2.7}), the non-linear parabolic eqs. (\ref{2.16}) for metals concentrations, the 
 semi-linear parabolic eqs. (\ref{2.11}) for substrate concentrations and the two 
 ordinary eqs. for the velocity of the microbial mass (\ref{2.2}) and the free boundary evolution (\ref{2.3}). 

 The complete model takes the following form:

			\[
			\frac{\partial X_i}{\partial t} +\frac{\partial}{\partial z}(u X_i)
			 =\rho_i r_{M,i}^{}(z,t,{\bf X},{\bf S},{\mbox{\boldmath $\mu$}}),\ i=1,...,n,\ 0\leq z\leq L(t),\ t>0,\\
			\]
\begin{equation}                                        \label{3.1}
       X_i(z,0)= X_{i0}(z),\ i=1,...,n,\ 0\leq z\leq L_0,\\
\end{equation}	
			
			\[
	     \frac{\partial u}{\partial z} = \sum_{i=1}^{n}r_{M,i}(z,t,\textbf{X},\textbf{S},{\mbox{\boldmath $\mu$}}),\ 0< z\leq L(t),\ t\geq0,\\ 
		  \]
\begin{equation}                                        \label{3.2}			
			u(0,t)=0,\  t\geq0,
\end{equation}
	
      \[
	    \dot L(t)=u(L(t),t)+\sigma_a(t)-\sigma_d(L(t)),\ t>0,
      \]
\begin{equation}                                        \label{3.3}
			L(0)= L_0,
\end{equation}		

\[
\frac{\partial \ \vartheta_i}{\partial t}
 +\frac{\partial}{\partial z}(u\vartheta_i)
 = r_{M,i}(z,t,{\bf X},{\bf S},{\mbox{\boldmath $\mu$}})- r_{D,i}(z,t,{\mbox{\boldmath $\mu$}},{\mbox{\boldmath $\vartheta$}},{\mbox{\boldmath $\vartheta$}})
 ,\ i=1,...,n,\ 0\leq z\leq L(t), t>0,\\
\]
\begin{equation}                                        \label{3.4}
	\vartheta_i(z,0)= \vartheta_{i0}(z),\ i=1,...,n, \  0\leq z\leq L_0,
\end{equation}		

\[
 \frac{\partial \ \bar{\vartheta_i}}{\partial t}
 +\frac{\partial}{\partial z}(u\bar{\vartheta_i})
 = r_{D,i}(z,t,{\mbox{\boldmath $\mu$}},{\mbox{\boldmath $\vartheta$}},{\mbox{\boldmath $\vartheta$}}),\ i=1,...,n,\ 0\leq z\leq L(t),\ t>0,\\
\]
\begin{equation}                                        \label{3.5}
\bar{\vartheta_i}(z,0)= \bar{\vartheta_{i0}}(z),\ i=1,...,n, \  0 < z < L_0, 
\end{equation}

\[
  \frac{\partial \mu_i}{\partial t}-\frac{\partial}{\partial z}\left(
   D_k\frac{\partial \mu_i}{\partial z}\right)=
   - Y_{ADS}N_ir_{D,i}(z,t,{\mbox{\boldmath $\mu$}},{\mbox{\boldmath $\vartheta_i$}},{\mbox{\boldmath $\bar{\vartheta_i}$}}),\ i=1,...,n,\ 0\leq z\leq L(t),\ t>0,\\
\]

 \begin{equation}                                        \label{3.6}
   \mu_i(z,0)=\mu_{i0}(z),\ 
   \frac{\partial \mu_i}{\partial z}(0,t)=0,\
   \mu_i(L(t),t)=\mu_{i L}(t),\ 0\leq z\leq L_0,\ t>0,\ i=1,...,n,
 \end{equation}

\[
	    \frac{\partial S_j}{\partial t}-\frac{\partial}{\partial z}\left(
   D_{S,j}\frac{\partial S_j}{\partial z}\right)=
   r_{S,j}(z,t,{\bf X},{\bf S},{\mbox{\boldmath $\mu$}}), \ j=1,...,m, 0 < z < L(t),\ t>0,
\]		  
\begin{equation}                                        \label{3.7}
  		S_j(z,0)=S_{j0}(z),\ \frac{\partial S_j}{\partial z}(0,t)=0,\ 
			S_j(L(t),t)=S_{jL},\ j=1,...,m, \ 0\leq z\leq L_0,\ t>0.
\end{equation}


 \noindent
 
 The index in system (\ref{3.1})-(\ref{3.7}) are specified as follows. Designate the active microbial species by the
 indexes $i=1,..,N$; inert materials by the index $i=N+1$; EPS by the index $i=N+2$.   

 The kinetic terms $r_{Mi}(z,t,\textbf{X},\textbf{S},{\mbox{\boldmath $\mu$}})$ in equation (\ref{3.1}) for the active
 microbial species $X_i$ can be expressed as follows:

 \begin{equation}                                        \label{3.8}
							r_{M,i} = (a_i-b_i-c_i)X_i \ \  i=1,...,N.
 \end{equation}		
 
 Inert biomass results from the natural decay of the active biomasses. Therefore the reaction term is expressed by

 \begin{equation}                                        \label{3.9}							
				 		  r_{M,N+1} = \sum_{i=1}^{N} c_{i} X_i\\.
 \end{equation}

 The formation rate of EPS resulting from the microbial metabolism is

\begin{equation}                                        \label{3.10}							
							r_{M,N+2} = \sum_{i=1}^{N} k_{i} a_i X_i. \\
\end{equation}

 The terms $a_i$ are the specific growth rate of the active biomass $X_i$ due
 to the uptake of substrates; $b_i$ represent the respiration rates for the
 single microbial species $X_i$; $c_i$ are the decay rates for the bacterial species.
 
 A coupled diffusion adsorption process has been considered for the contaminant
 $\mu_i$. A reversible mechanism for bio sorption has been 
 considered. The sorbent contaminant adsorption-desorption terms are

 \begin{equation}                                        \label{3.11}
   r_{D,i}= k_{ADS}\mu\vartheta_i - k_{DES}\bar{\vartheta_i} 
 \end{equation}

 where $k_{ADS,i}$ denotes the adsorption constant, $k_{DES,i}$ denotes the desorption constant. 


 \vspace{5mm}\noindent
\noindent
 \textbf{Remark 1}.

\noindent
Let us consider a biofilm characterized by no-growth condition, by the absence of attachment and detachment phenomena and by the presence of a single sorbent contaminant $\mu$ wich adsorb equally on all the different biofilm components.
Under these condition the growth velocity $u(z,t)$ of the microbial biomass, the attachment biomass flux $\sigma_a^{}(t)$ and the detachment biomass flux $\sigma_d^{}(L(t),t)$ are equal to $0$. 

Considering $u(z,t)=0$ in Eqs. (\ref{2.12})


\begin{equation}                                        \label{3.13}
 \frac{\partial \ \bar{\vartheta_i}}{\partial t}
 = r_{D,i}(z,t,{\mbox{\boldmath $\mu$}},{\mbox{\boldmath $\vartheta$}}),
 \end{equation}

 with

 \begin{equation}                                        \label{3.14}
   r_{D,i}= k_{ADS}\mu\vartheta - k_{DES}\bar{\vartheta}.
 \end{equation}
 
 Summing (\ref{3.13}) on $i$ yields

 \begin{equation}                                        \label{3.15}
  \frac{\partial \ \Theta}{\partial t}
 = K_{ads} \mu (1-\Theta) - K_{des} \Theta
 \end{equation}

 where
 \[
 \Theta = \sum_{i=1}^{n} \bar{\vartheta_i}
 \]
 \[
 1-\Theta = \sum_{i=1}^{n} \vartheta_i
 \]
 
Eq. \ref{3.15} is the coupled diffusion-adsorption approach for thin film firstly
introduced by Bartlett and Gardner \cite{bartlett1996diffusion}. 
Thus the model presented in \cite{bartlett1996diffusion,d2015mathematical} can be considered
 a special case of the model presented in this work.



\section{Mathematical modeling of real Systems}\label{n4}
 \setcounter{equation}{0} 

  This section is devoted to the mathematical modelling of two special cases of particular
 biological and engineering interest. The two
 models presented refer to special biofilm systems interacting with sorbent
 contaminants. For each system analyzed, model equations have been defined and numerical simulations have been reported.
 In the first case (Case 1),  the dynamics of
 a single sorbent contaminant have been modeled. In particular, the effect of binding
 site density (Case 1a) and the effect of different adsorption rates (Case 1b) have been
 analyzed. In the second case (Case 2), the same biofilm configuration system of case 1 has
 been assumed. However the dynamics of two different contaminants including their sorption
 on two different biofilm components have been analyzed. 


For all the systems analyzed, numerical solutions to the free
 boundary problem stated in Section $\ref{n3}$ have been obtained by using the method
 of characteristics, e.g. \cite{art:rif.17,art:rif.18}. Accuracy was checked by comparison to
 the geometric constraint $\sum_{i=1}^{n}f_i(z,t)=1$. Simulations were performed using original software
 developed for this work.

\subsection{Case 1}\label{n4.1}
 A heterotrophic-autotrophic competition biofilm system, for ammonia nitrification and
 organic carbon degradation has been considered. This biofilm system is typical of a urban wastewater treatment plant.
 EPS production is also taken into account following the approach proposed by \cite{laspidou2002non}.
 The biomass increase is determined by the metabolism of the dissolved components. 
 The model considers the growth and decay of four different biofilm components, including
 heterotrophic bacteria $X_1=\rho_1f_1$, autotrophic bacteria $X_2=\rho_2f_2$, inert material
 $X_3=\rho_3f_3$ and EPS $X_4=\rho_4f_4$. Three substrate, ammonia $S_1$, Organic carbon $S_2$, 
 and Oxygen $S_3$ are taken into account.  
 The autotrophs are nitrifying bacteria that grow by consuming ammonia and oxygen. The heterotrophic bacteria
 uptake organic carbon and oxygen. The two species compete for space and oxygen \cite{art:rif.8}.
 The active biomass $X_1$ and $X_2$ is consumed via respiration and decay processes, producing residual inert microbial
 biomass $X_3 = X_{In}$.

 \begin{figure*}
 \includegraphics[width=0.6\textwidth]{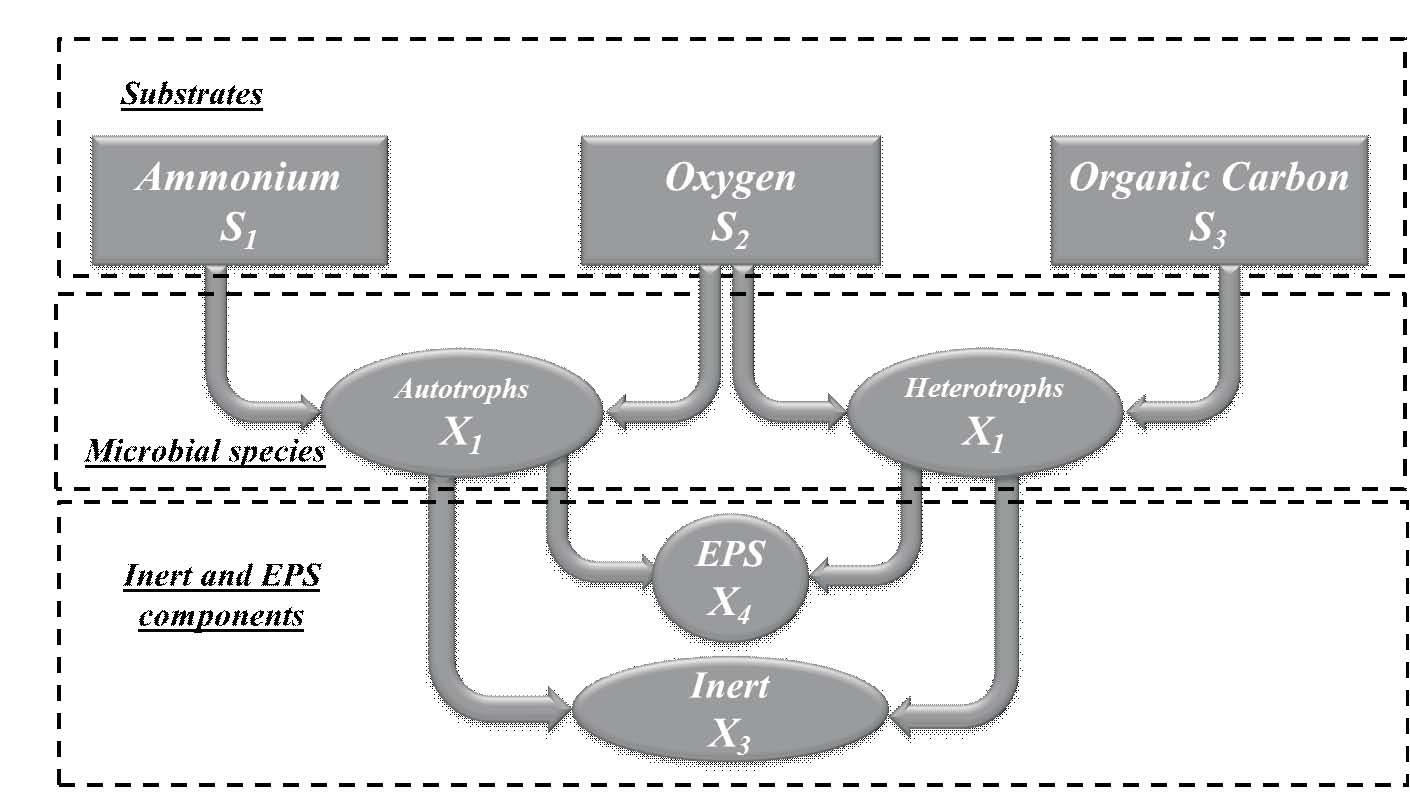}
 \caption{Schematic representation of kinetic process.}
          \label{f4.1a}     
 \end{figure*}

 In this first case, one contaminant has been taken into account and its interactions
 with biofilm matrix components have been studied. In Case 1a four different binding site densities have
 been considered. The related numerical simulations have been run to asses the influence of
 sites density on the diffusion of the contaminants within the biofilm and thus on the adsorption
 phenomenon evolution. In the Case 1b, four different adsorption kinetic constant values have been tested.
 
 For the specific system analyzed here, the kinetic terms $r_{Mi}(z,t,\textbf{X},\textbf{S},
 {\mbox{\boldmath $\mu$}})$ in equation (\ref{3.1}) for the biofilm components
  $X_1$, $X_2$, $X_3$, and $X_4$ can be expressed as follows:

 \begin{equation}                                        \label{4.2}
							r_{M,1} = (a_1-b_1-c_1)X_1, \\
 \end{equation}		
 
 \begin{equation}                                        \label{4.3}
							r_{M,2} = (a_2-b_2-c_2)X_2, \\
 \end{equation}

 For the Inert components $X_3$

 \begin{equation}                                        \label{4.4}
							r_{M,3} = c_1X_1+c_2X_2, \\
 \end{equation}

 while for EPS components $X_4$

 \begin{equation}                                        \label{4.5}
							r_{M,4} = k_1a_1+k_2a_2. \\
 \end{equation}

 The terms $a_1$ and $a_2$ are the specific growth rate of the active biomass $X_1$ and $X_2$ due to the uptake of substrates; 
 $b_1$ and $b_2$ represent the respiration rates for the single microbial species $X_1$ and $X_2$, respectively.
 $c_1$ and $c_2$ are the decay rates for the heterotrophic ($X_1$) and autotrophic ($X_2$) microorganisms.
 
 They are given by:
 
 \begin{equation}                                        \label{4.6}
 a_1^{}=(1-k_1)K_{\max,1}\frac{S_2}{K_{1,2}+S_2}
       \frac{S_3}{K_{1,3}+S_3},
 \end{equation}
 \begin{equation}                                        \label{4.7}
 a_2^{}=(1-k_2)K_{\max,2}
       \frac{S_1}{K_{2,1}+S_1}
 			 \frac{S_3}{K_{2,3}+S_3},
 \end{equation}
 \begin{equation}                                        \label{4.8}
  b_1^{}=
         b_{m,1}F_1\frac{S_3}{K_{1,3}+S_3},
 \end{equation}
 \begin{equation}                                        \label{4.9}
  b_2^{}=
         b_{m,2}F_2\frac{S_3}{K_{2,3}+S_3},
 \end{equation}
 \begin{equation}                                        \label{4.10}
  c_1^{}= 
        (1-F_1)c_{m,1},
 \end{equation}
\begin{equation}                                         \label{4.11}
  c_2^{}= 
       (1-F_2)c_{m,2},
 \end{equation}
 
where $\mu_{\max,i}$ denotes the maximum net growth rate for biomass $i$, $k_i^{}$ is the the 
growth-associated EPS formation coefficient, $K_{i,j}$ 
the affinity constant of substrate $j$ for biomass $i$, $b_{m,i}$ the
endogenous rate for biomass $i$, $c_{m,i}$ the
decay-inactivation rate for biomass $i$, $F_i^{}$ the biodegradable fraction of biomass $i$. 

The conversion rates of substrates $r_{S,j}^{}(z,t,{\bf X},{\bf S})$ in equation (\ref{3.7})
are expressed by:

\begin{equation}                                        \label{4.12}
r_{{S,1}^{}} = -\frac{1}{Y_2}a_2^{}X_2,
\end{equation}
\begin{equation}                                        \label{4.13}
r_{{S,2}^{}} =  -\frac{1}{Y_1}a_1^{}X_1,
\end{equation}
\begin{equation}                                        \label{4.14}
r_{{S,3}^{}}= -(1-k_1)\frac{(1-Y_1)}{Y_1}a_1^{}X_1 -(1-k_2)\frac{(1-Y_2)}{Y_2}a_2^{}X_2
							- b_1^{}X_1 - b_2^{}X_2,
\end{equation}

where $Y_i$ denotes the yield for biomass $i$. 

 A coupled diffusion adsorption process has been considered for a sorbent component
 $\mu_1= \mu $. A non-reversible mechanism for contaminant sorption has been 
 considered. The sorbent contaminant adsorption term is

 \begin{equation}                                        \label{4.15}
   r_{D,1}= K_{ads}\mu\vartheta_1
 \end{equation}

\begin{table}[ht]
\begin{footnotesize}
 \begin{center}
 \begin{tabular}{llccc}
 \hline
{\textbf{Parameter}} & {\textbf{Definition}} & {\textbf{Unit}} &  \textbf{Value} &  \textbf{References} \\
 \hline
 $\mu_{max_1}$   & Maximum growth rate for $X_1$              &  $d^{-1}$                              & 4.8       & \cite{art:rif.8}\\
 $\mu_{max_2}$   & Maximum growth rate for $X_2$              &  $d^{-1}$                              & 0.95      & \cite{art:rif.8}\\
 $K_1$         & EPS formation by $X_1$                       &  $mgCOD/mgCOD$                         & 0.02      & adapted from \cite{merkey2009modeling}\\
 $K_2$         & EPS formation by $X_2$                       &  $mgCOD/mgCOD$                         & 0.011     & adapted from \cite{merkey2009modeling}\\
 $Ks_{1,2}$    & Organics half saturation constant for $X_1$  &  $mgCODl^{-1}$                         & 5         & \cite{art:rif.8}\\
 $Ks_{1,3}$    & Oxygen half saturation constant for $X_1$    &  $mgl^{-1}$                            & 0.1       & \cite{art:rif.8}\\
 $Ks_{2,1}$    & Ammonium half saturation constant for $X_2$  &  $mgNl^{-1}$                           & 1         & \cite{art:rif.8}\\
 $Ks_{2,3}$    & Oxygen half saturation constant for $X_2$    &  $mgl^{-1}$                            & 0.1       & \cite{art:rif.8}\\
 $b_{m,1}$     & Endogenous rate for $X_1$                    &  $d^{-1}$                              & 0.025     & \cite{merkey2009modeling}\\
 $b_{m,2}$     & Endogenous rate for $X_2$                    &  $d^{-1}$                              & 0.0625    & \cite{merkey2009modeling}\\
 $F_{1}$       & Biodegradable fraction of $X_1$              &  $--$                                  & 0.8       & \cite{merkey2009modeling}\\
 $F_{2}$       & Biodegradable fraction of $X_2$              &  $--$                                  & 0.8       & \cite{merkey2009modeling}\\
 $c_{m,1}$     & Decay-inactivation rate for $X_1$            &  $d^{-1}$                              & 0.05      & \cite{merkey2009modeling}\\
 $c_{m,2}$     & Decay-inactivation rate for $X_2$            &  $d^{-1}$                              & 0.05      & \cite{merkey2009modeling}\\
 $Y_1$         & Yield of $X_1$                               &  ${g_{biomass}}/{g_{substrate}}$       & 0.4       & \cite{art:rif.8}\\ 
 $Y_2$         & Yield of $X_2$                               &  ${g_{biomass}}/{g_{substrate}}$       & 0.22      & \cite{art:rif.8}\\
 $Y_{ads}$       & Yield of adsorbent                           &  $--$                                  & 1                               & This study\\ 
 $K_{ads,1}$   & Adsorption kinetic constant for $X_1$        &  $d^{-1}$                              & $5\cdot10^{3}$,$5\cdot10^{2}$,$5\cdot10$,$5$  &This study\\ 		
 $N_{b,1}$	   & Bounding sites density for $X_1$ 				    &  $mgl^{-1}$                            & 2,10,100,1000              &This study\\ 
 $K_{I1}$      & Inhibition constant                          &  $h^{-1}$                              & 4.17      & This study\\
 $\rho$        & biofilm density                                    &  $gm^{-3}$                          & 65000  & This study\\
 $\lambda$     & Biomass shear constant                             &  $mmh^{-1}$                         & 2000   & This study\\
 \hline
 \end{tabular}
 \caption{Kinetic parameters used for model simulations} \label{t4.1}
 \end{center}
 \end{footnotesize}
 \end{table}

\begin{table}[hb]
\begin{footnotesize}
 \begin{center}
 \begin{tabular}{lclcc}
 \hline
{\textbf{Parameter}} & {\textbf{Symbol}} & {\textbf{Unit}} &  \textbf{Value} \\
 \hline
 
 COD concentration    at $L=L(t)$             &  $S_{1L}$       &  mg$l^{-1}$       & 20  \\
 Oxygen concentration  at $L=L(t)$            &  $S_{3L}$       &  mg$l^{-1}$       & 8     \\
 Ammonium concentration  at $L=L(t)$          &  $S_{2L}$       &  mg$l^{-1}$       & 2     \\
 Time Simulation                              &  T              &  d                & 100   \\
 Initial Biofilm thickness                    &  $L_0$          & mm                & 0.3   \\
 Initial Volume Fraction of Autotrophs $(X_1)$        &  $f_{1,0}(z)$   & --                & 0.399  \\
 Initial Volume Fraction of Heterotrophs $(X_2)$      &  $f_{2,0}(z)$   & --                & 0.5   \\
 Initial Volume Fraction of Inert        $(X_3)$      &  $f_{3,0}(z)$   & --                & 0.001   \\
 Initial Volume Fraction of EPS          $(X_4)$      &  $f_{5,0}(z)$   & --                & 0.1   \\

 \hline
 \end{tabular}
 \caption{Initial conditions for biofilm growth} \label{t4.2}
 \end{center}
 \end{footnotesize}
 \end{table}
 
\noindent

 For all the dissolved species, substrates and sorbent contaminant, Dirichlet condition on the free boundary
 have been assumed. In eq. (\ref{3.3}) governing the free boundary evolution, $\sigma_d(L(t))$ 
 is assumed to be a known function of $L$ and $t$: 
 
 \begin{equation}                                        \label{4.16}
   \sigma_d(L(t))= \lambda L_{}^{2}(t) 
 \end{equation}

 where $\lambda$ is the share constant whose value is reported in Table \ref{t4.1}. 
 No attachment phenomena have been considered for all the simulation performed, thus $\sigma_a(t)$ has
 been fixed to zero. The stoichiometry and kinetic parameter values used in the model 
 are reported in Table \ref{t4.1}.
 The model outputs are reported in figures \ref{f4.1}-\ref{f4.7}. Numerical simulations demonstrate model capability 
 of predicting biofilm component distributions, occupied and free binding sites fractions, substrate trends, free
 contaminants profiles over biofilm depth and biofilm thickness.
 The initial biofilm composition has been defined in Table \ref{t4.2}. In particular, the biofilm is set to be
 initially constituted by Autotrophic component (39.9\%), Heterotrophic component (50\%), EPS (10\%) and Inert (0.1\%)
 with an initial biofilm thickness of 300 $\mu m$.
 The simulations reproduce the environmental conditions in a wastewater treatment plant. The oxygen 
 concentration in the bulk liquid has been fixed to 8 mg/l, consistent with continuous aerated systems.
 The concentrations of soluble COD and ammonium in the bulk liquid are fixed on 20 mg/l and 2 mg/l, respectively.

\subsubsection{Case 1a: Effect of Site density on adsorption phenomenon}\label{n4.1.1}

\begin{figure*}
 \includegraphics[width=1\textwidth]{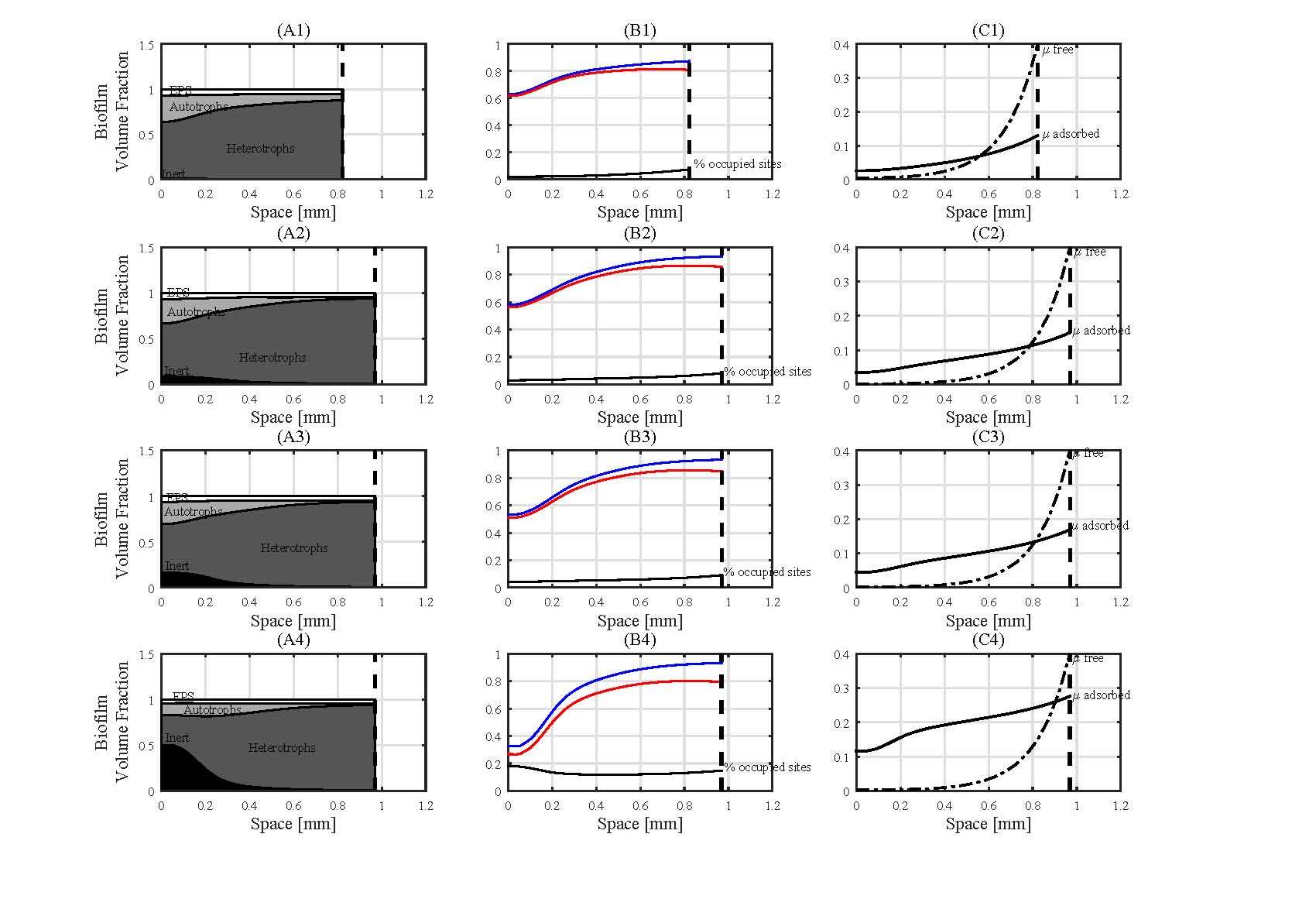}
 \caption{Effect of Site density $N_b=2$ on adsorption phenomenon. Microbial species distribution (A1-A4); total (blue-line) binding sites fraction, free binding sites fractions (red-line) percentage of occupied sites (B1-B4); adsorbed and free metal profile (C1-C4) after 1 (A1,B1,C1), 10(A2,B2,C2), 20(A3,B3,C3), 100(A4,B4,C4) days simulation time. Free contaminant concentration is multiplied by a factor of $10^4$.}
          \label{f4.1}
 \end{figure*}

 Four different values of binding site density $N_b$ are used in numerical simulations. 
 Figure \ref{f4.1} shows the evolution of biofilm distribution (A1-A4), total and free binding sites fractions,
 percentage of occupied sites (B1-B4) and adsorbed and free metal (C1-C4) after 1(A1,B1,C1), 10(A2,B2,C2),
 20(A3,B3,C3), 100(A4,B4,C4) days simulation time. In the first numerical simulation a binding site density $N_b$ equal to 2
 is considered. After 1 day the microbial distribution resembles already a typical
 heterotrophic autotrophic stratification, with the heterotrophic bacteria dominating the outmost part of the
 biofilm (fig. \ref{f4.1} A1). Free contaminant concentration (fig. \ref{f4.1} C1-C4) shows a parabolic trend 
 decreasing from the right side (bulk liquid) to the left side (substratum). As expected, the accumulation of
 sorbed contaminant is higher in the external part of biofilm than in the inner part, in agreement with the
 the adsorption rate dependance on the concentration of the free contaminants (eq.\ref{4.14}).
 The percentage of occupied sites is slightly increasing for the first 20 days of time simulation (fig. \ref{f4.1} B1--B3). 
 After 20 days the increase of inert fraction, in the inner part, reduce the fraction of heterotrophic bacteria fraction
 and then the total amount of the relative free binding sites. This is the reason why the percentage of occupied binding
 sites is higher than in the external part of the biofilm.

\begin{figure*}
 \includegraphics[width=1\textwidth]{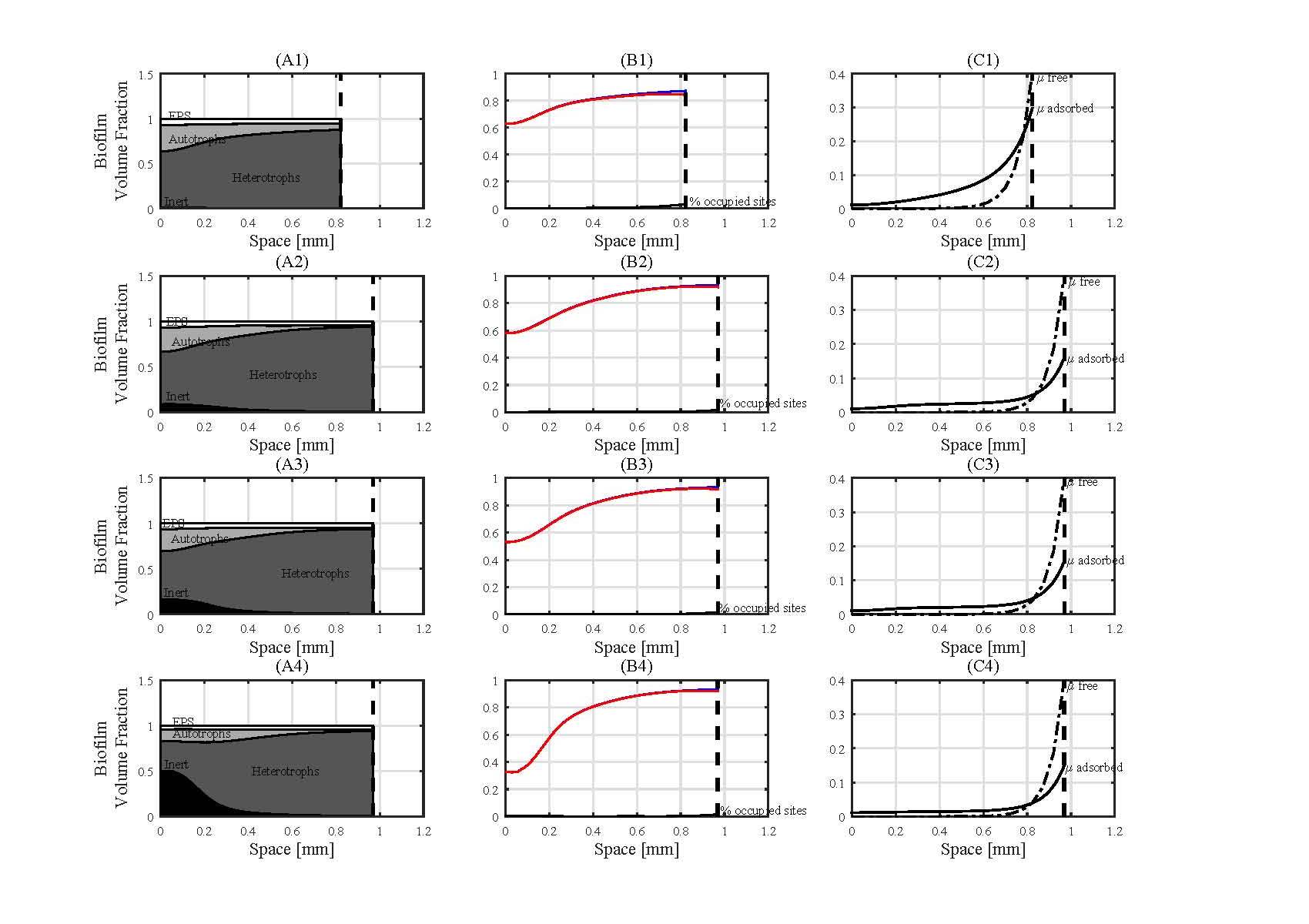}
 \caption{Effect of Site density $N_b=10$ on adsorption phenomenon. Microbial species distribution (A1-A4); total (blue-line) binding sites fraction, free binding sites fractions (red-line) percentage of occupied sites (B1-B4); adsorbed and free metal profile (C1-C4) after 1 (A1,B1,C1), 10(A2,B2,C2), 20(A3,B3,C3), 100(A4,B4,C4) days simulation time. Free contaminant concentration is multiplied by a factor of $10^4$.}
          \label{f4.2}
  \end{figure*}

\begin{figure*}
 \includegraphics[width=1\textwidth]{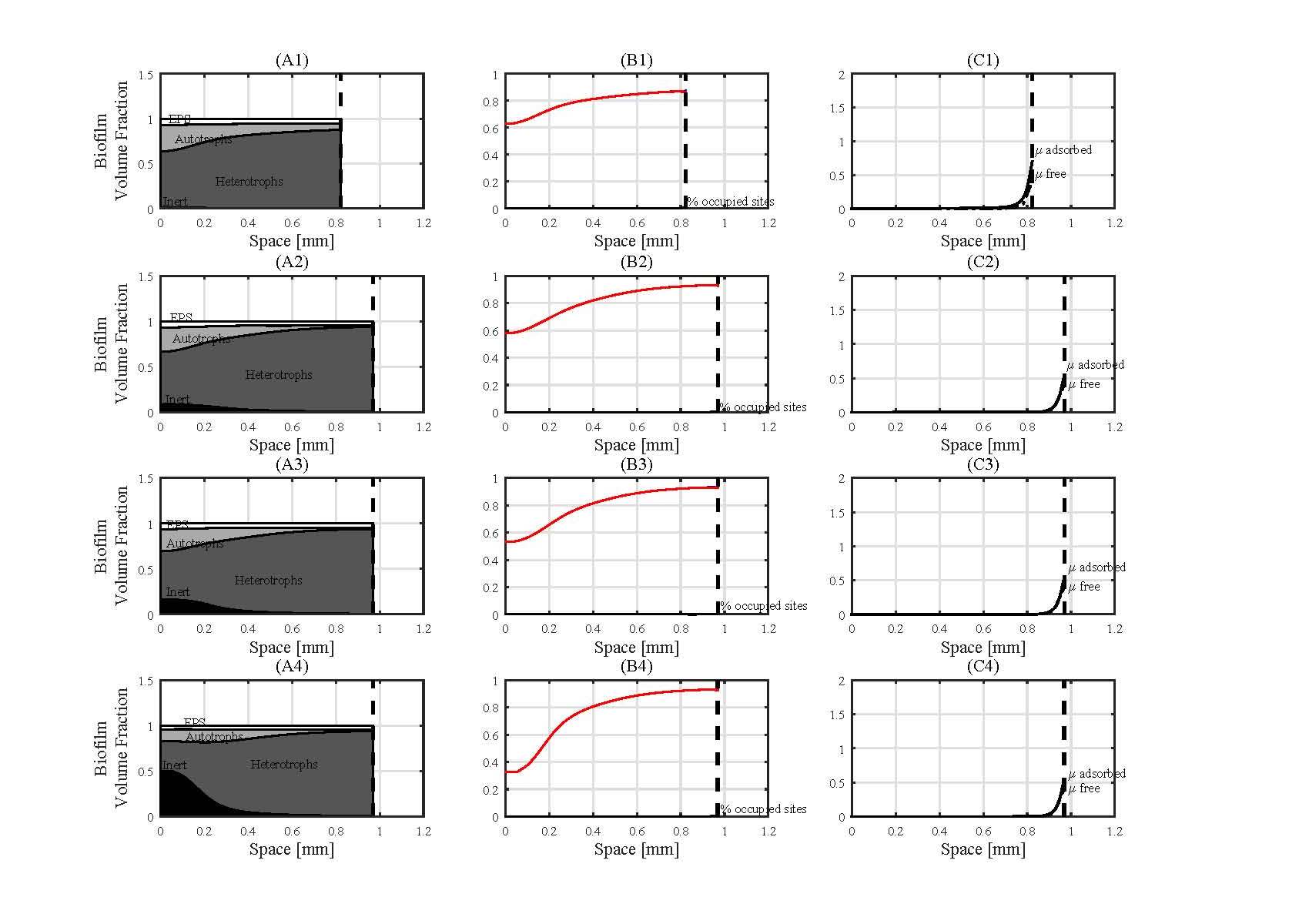}
 \caption{Effect of Site density $N_b=100$ on adsorption phenomenon. Microbial species distribution (A1-A4); total (blue-line) binding sites fraction, free binding sites fractions (red-line) percentage of occupied sites (B1-B4); adsorbed and free metal profile (C1-C4) after 1 (A1,B1,C1), 10(A2,B2,C2), 20(A3,B3,C3), 100(A4,B4,C4) days simulation time. Free contaminant concentration is multiplied by a factor of $10^4$.}
          \label{f4.3}
  \end{figure*}

\begin{figure*}
 \includegraphics[width=1\textwidth]{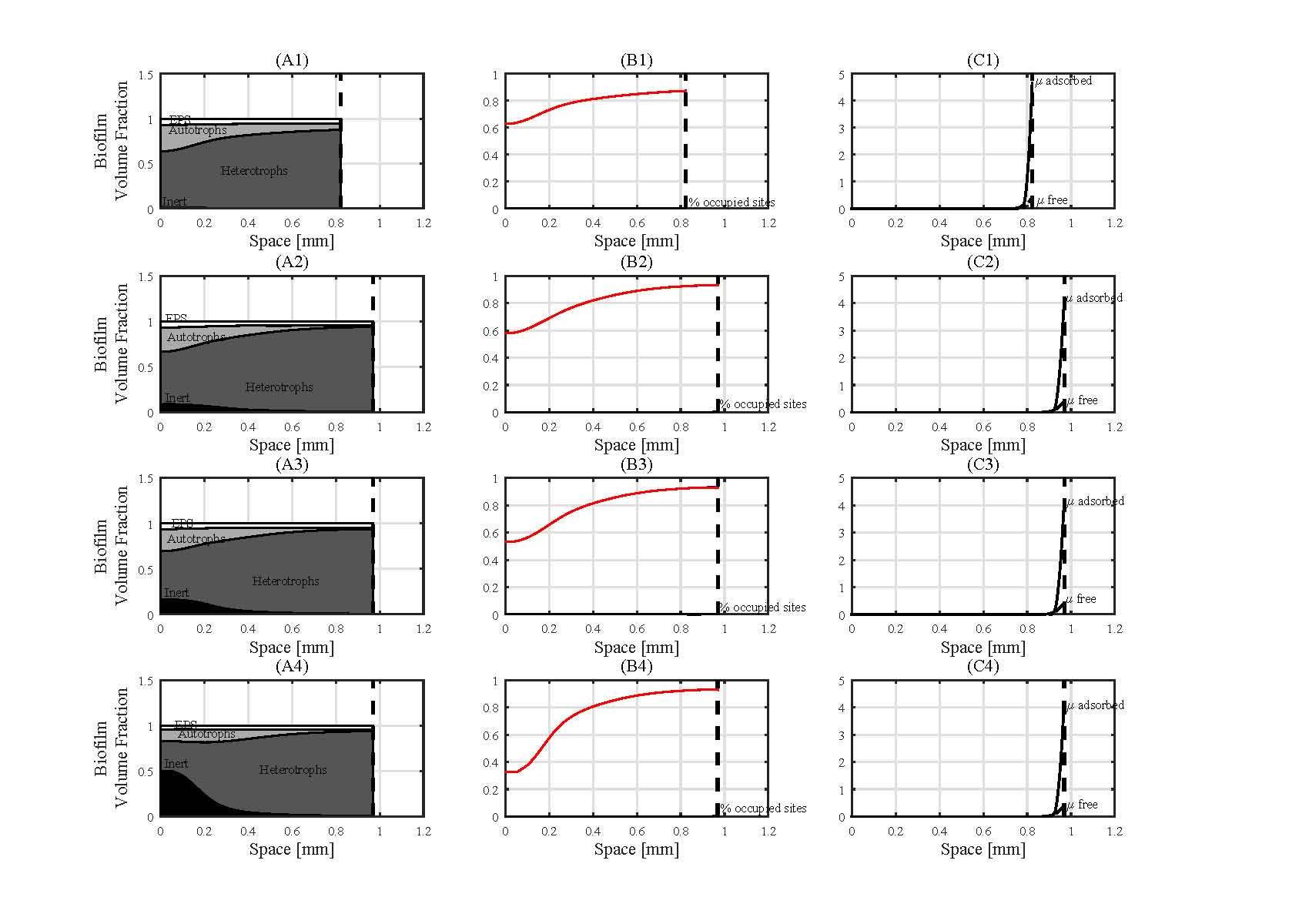}
 \caption{Effect of Site density $N_b=1000$ on adsorption phenomenon. Microbial species distribution (A1-A4); total (blue-line) binding sites fraction, free binding sites fractions (red-line) percentage of occupied sites (B1-B4); adsorbed and free metal profile (C1-C4) after 1 (A1,B1,C1), 10(A2,B2,C2), 20(A3,B3,C3), 100(A4,B4,C4) days simulation time. Free contaminant concentration is multiplied by a factor of $10^4$.}
          \label{f4.4}
  \end{figure*}

 Figures \ref{f4.2} shows the evolution of the considered biological system, for the binding
 site density $N_b=10$. Differently from the previous simulation, the biofilm is not fully penetrated by
 the free contaminant (fig. \ref{f4.1} C1--C4). This determines the accumulation of the sorbed contaminant
 only in the external layer of the biofilm (fig. \ref{f4.1} B1--B4). Due to the higher binding site density
 the percentage of the occupied binding sites is almost equal to $1\%$.
 It is interesting to note that, the concentration of sorbed contaminants on the free boundary after 20
 days simulation time is constant for all the four different simulations (fig. \ref{f4.1}--\ref{f4.4} C2--C4). 
 This occurs because, due to the shear stress effect, the superficial layer of the biofilm and thus the
 adsorbed contaminant is continuously removed. By considering that the shear stress has been modeled as a
 function of $L$ (\ref{4.16}), when the biofilm thickness is stable (in the figures \ref{f4.1}--\ref{f4.4}
 this is visible after 10 days) the shear stress becomes constant. Accordingly, the equilibrium between the
 contaminant that absorbs on the biofilm and the sorbed contaminant removed from the shear stress has been reached. 

In figures \ref{f4.3} and \ref{f4.4} the binding site densities are set to 100 and 1000 respectively. In these two simulations
it is possible to note that the adsorption phenomenon occurs only in the external part of the biofilm. Even if not visible in the figure, in the outmost layer of the biofilm the percentage of occupied sites differs from zero. Due to the higher binding site availability, the total amount of adsorbed contaminant on the free boundary increases with the increase of the binding site density
(fig. \ref{f4.3} C1--C4 and \ref{f4.4} C1--C4).

 \subsubsection{Case 1b: Effect of adsorption rate}\label{n4.1.2}

\begin{figure*}
 \includegraphics[width=1\textwidth]{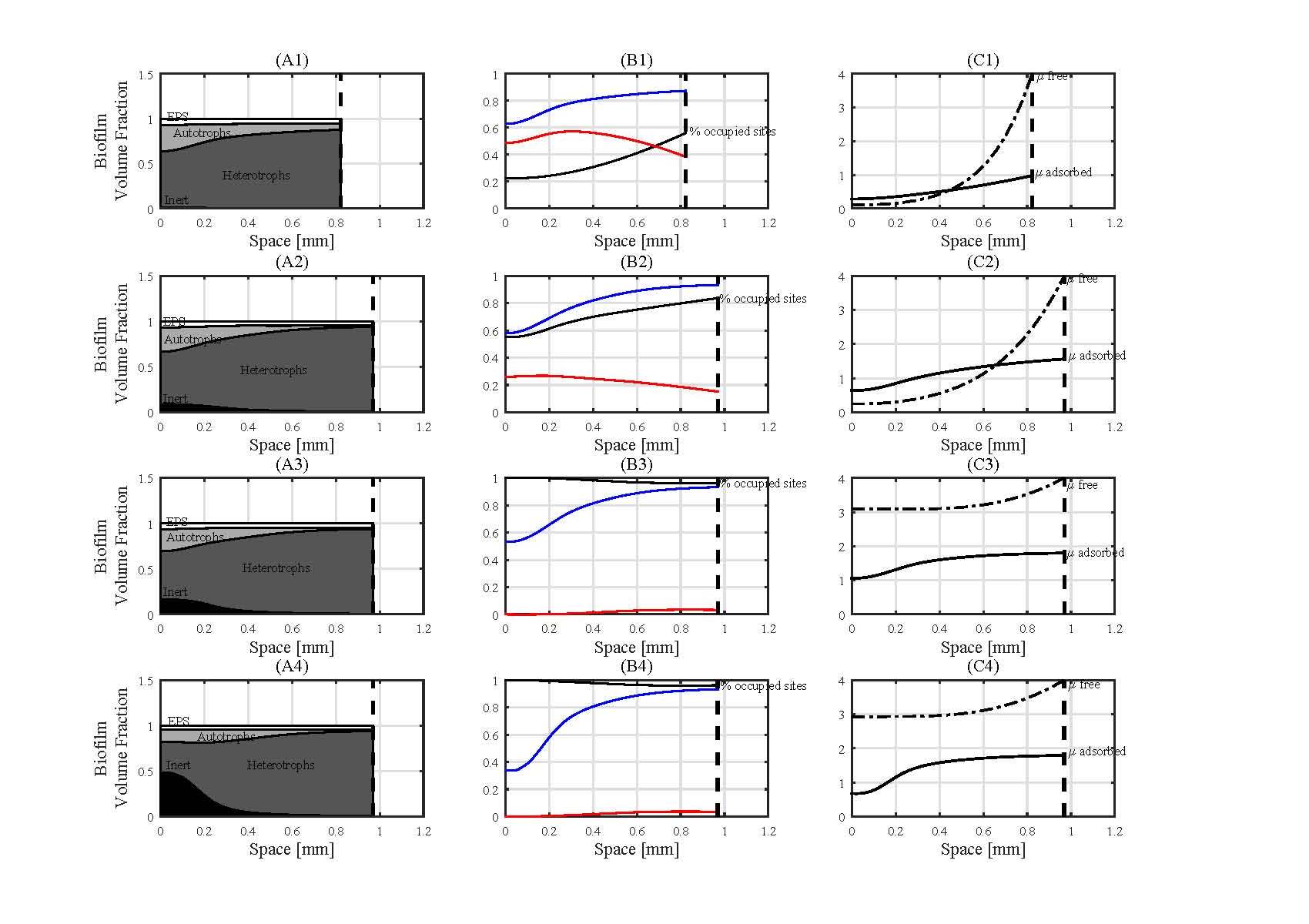}
 \caption{Effect of adsorption constant $K_{ads,1}=5\cdot10^3$ with $N_b=2$ on adsorption phenomenon. Microbial species distribution (A1-A4); total (blue-line) binding sites fraction, free binding sites fractions (red-line) percentage of occupied sites (B1-B4); adsorbed and free metal profile (C1-C4) after 1 (A1,B1,C1), 10(A2,B2,C2), 20(A3,B3,C3), 100(A4,B4,C4) days simulation time. Free contaminant concentration is multiplied by a factor of $10^4$.}
          \label{f4.5}
  \end{figure*}

\begin{figure*}
 \includegraphics[width=1\textwidth]{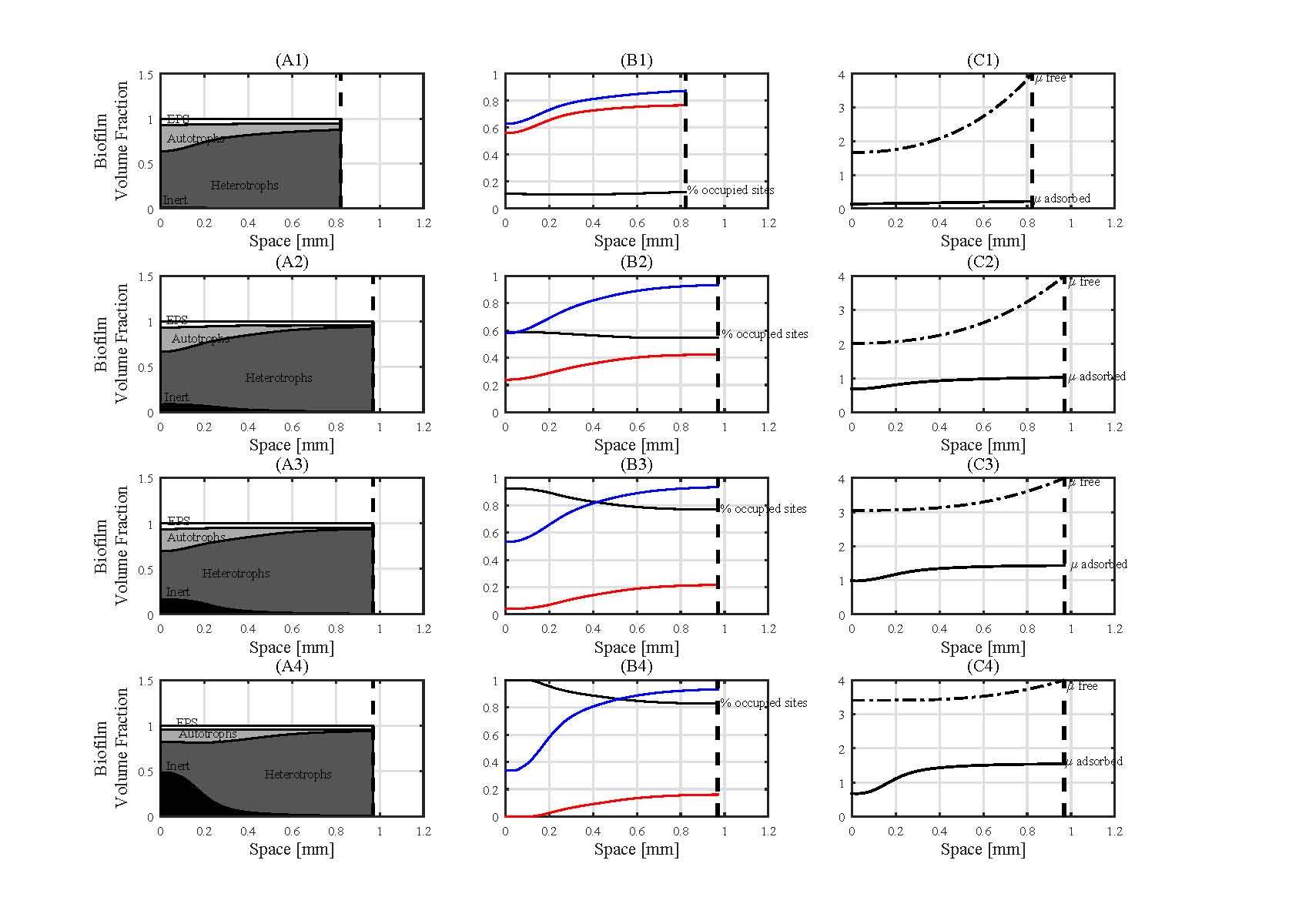}
 \caption{Effect of adsorption constant $K_{ads,1}=5\cdot10^2$ with $N_b=2$ on adsorption phenomenon. Microbial species distribution (A1-A4); total (blue-line) binding sites fraction, free binding sites fractions (red-line) percentage of occupied sites (B1-B4); adsorbed and free metal profile (C1-C4) after 1 (A1,B1,C1), 10(A2,B2,C2), 20(A3,B3,C3), 100(A4,B4,C4) days simulation time. Free contaminant concentration is multiplied by a factor of $10^4$.}
          \label{f4.6}
 \end{figure*}

\begin{figure*}
 \includegraphics[width=1\textwidth]{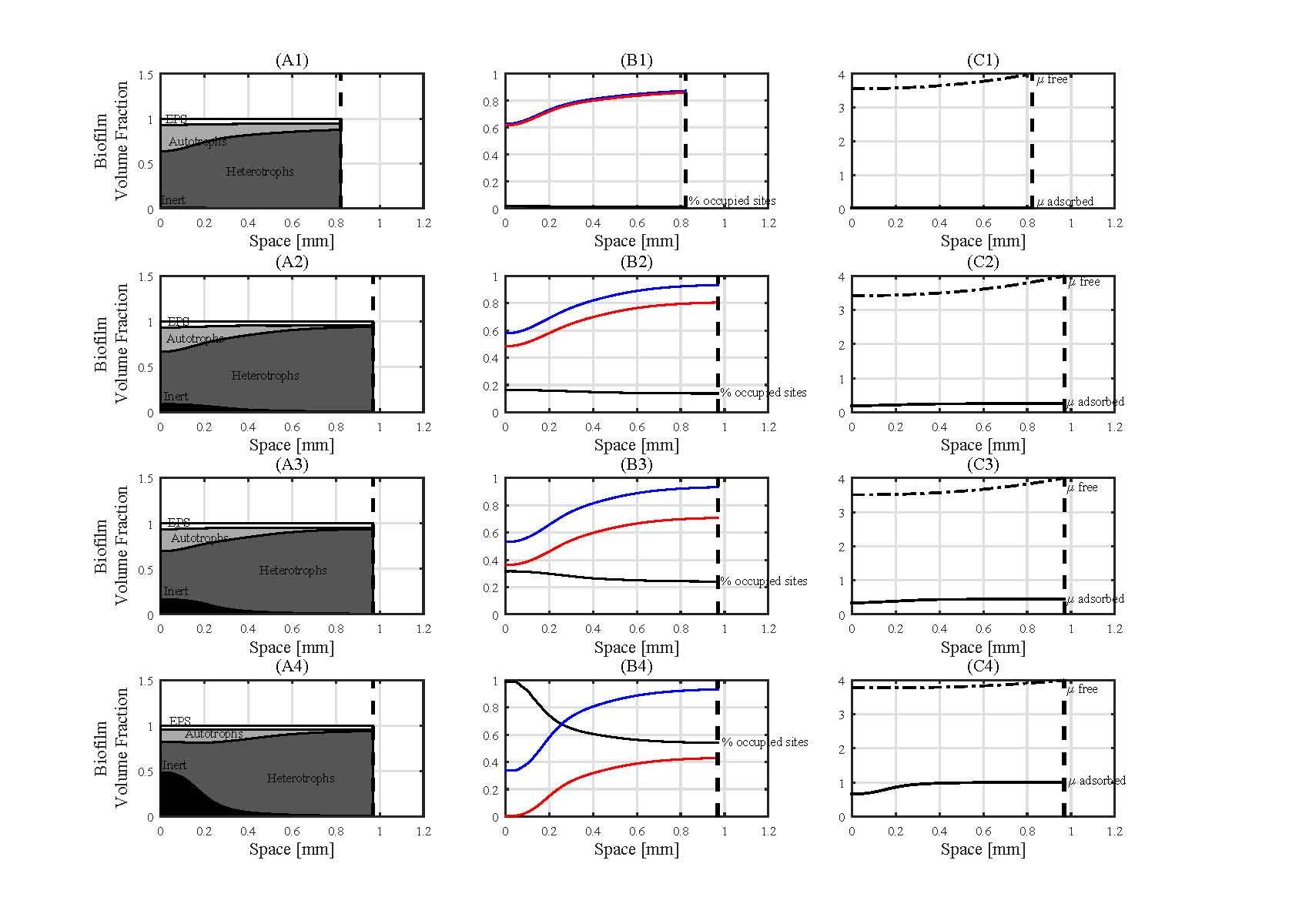}
 \caption{Effect of adsorption constant $K_{ads,1}=50$ with $N_b=2$ on adsorption phenomenon. Microbial species distribution (A1-A4); total (blue-line) binding sites fraction, free binding sites fractions (red-line) percentage of occupied sites (B1-B4); adsorbed and free metal profile (C1-C4) after 1 (A1,B1,C1), 10(A2,B2,C2), 20(A3,B3,C3), 100(A4,B4,C4) days simulation time. Free contaminant concentration is multiplied by a factor of $10^4$.}
          \label{f4.7}
  \end{figure*}

 In these simulations different values of the adsorption rate constant $K_{ads,1}$ have been taken into account 
 (fig. \ref{f4.5}--\ref{f4.7}), while the value of binding site density $N_b$ has been fixed equal to 2. As in the previous  
 simulation set, the heterotrophic bacteria predominate the central and the external part of the biofilm, with the autotrophic   
 bacteria are present in the inner layer (fig. \ref{f4.5}--\ref{f4.7} A1--A4).
 In figure \ref{f4.5} the contaminant bio-sorbs on a multispecies biofilm system with a sorption rate constant equal to 
 $K_{ads,1}=5\cdot10^3$. The biofilm is fully penetrated by the free contaminant and the entire biofilm is affected by
 the sorption process. According to this, after 1 day simulation time, the concentration of the sorbed contaminants and
 the percentage of the occupied sites are different from zero (fig. \ref{f4.5} B1 and C1).
 Going on with the simulation time, the accumulation of contaminant continues where free binding sites are present. According
 to this the fraction of occupied binding sites increases over the time and reaches a stable value after 20 days (fig. \ref{f4.5} B3).
 
 It is interesting to note that the fraction of occupied sites is equal to $100\%$ only in the inner part of the biofilm
 where there is a lack of substrate (simulations not shown), the growth velocity of the biofilm is very low, and thus the velocity of formation of
 new binding sites is less than the sorption velocity
 (fig. \ref{f4.5} B3 and B4). In the external layer of biofilm the substrates are abundant (simulations not shown) 
 and the velocity of the formation of new binding sites is higher than the sorption velocity, for this reason the biofilm
 is not fully saturated and the equilibrium is reached. As it is possible to see after 20 days simulation time the
 profile of occupied sites does not change.

 Figure \ref{f4.6} shows the simulation results for the considered biological system with a $K_{ads,1}=5\cdot10^2$. Simulation results confirm that lower value
 of the adsorption kinetic constant determines a slow adsorption rate, as shown as well in fig. \ref{f4.7}.

 \subsection{Case 2: HMs fractionation in biofilm components}\label{n4.2}

 In this special case the fate of two different contaminants that bio-sorb on two different biofilm components
 has been considered. The same biological system constituted by a heterotrophic autotrophic biofilm described
 in the previous section has been modelled.
 In particular, the same biofilm components, $X_i,i=1,...,4$ and the same substrates $S_j,j=1,2$ as in the previous
 case are taken into account. The two contaminants $\mu_1$ and $\mu_2$ are considered to adsorb on heterotrophic biomass $X_1$ and Inert $X_3$
 respectively. The different \textit{\textit{affinity}} of the two contaminants to the different biofilm 
 components follows from experimental observation. Indeed the different nature (e.g \textit{gram positive} or
 \textit{gram negative}) of the microbial species can affect the characteristics of the cell walls in terms of
 binding sites \textit{affinity}. Moreover the non-microbial biofilm components, such as Inert and EPS, show very
 high sorbent capacities.
 The system is governed by equations (\ref{4.2})--(\ref{4.14}). In particular, the diffusion and the reaction of the two
 sorbent contaminats $\mu_1$ and $\mu_2$ have been modeled considering a non-reversible mechanism
 The contaminant adsorption terms are

 \begin{equation}                                        \label{4.17}
   r_{D,1}= k_{ads}\mu_1\vartheta_1
 \end{equation}

 \begin{equation}                                        \label{4.18}
   r_{D,2}= k_{ads}\mu_1\vartheta_ 3
 \end{equation}

The same boundary conditions of the previous application have
 been considered, except for $\mu_2$ that requests a new one: $\mu_2(L(t),t)=4*10^{-4}$.

\begin{figure*}
 \includegraphics[width=1\textwidth]{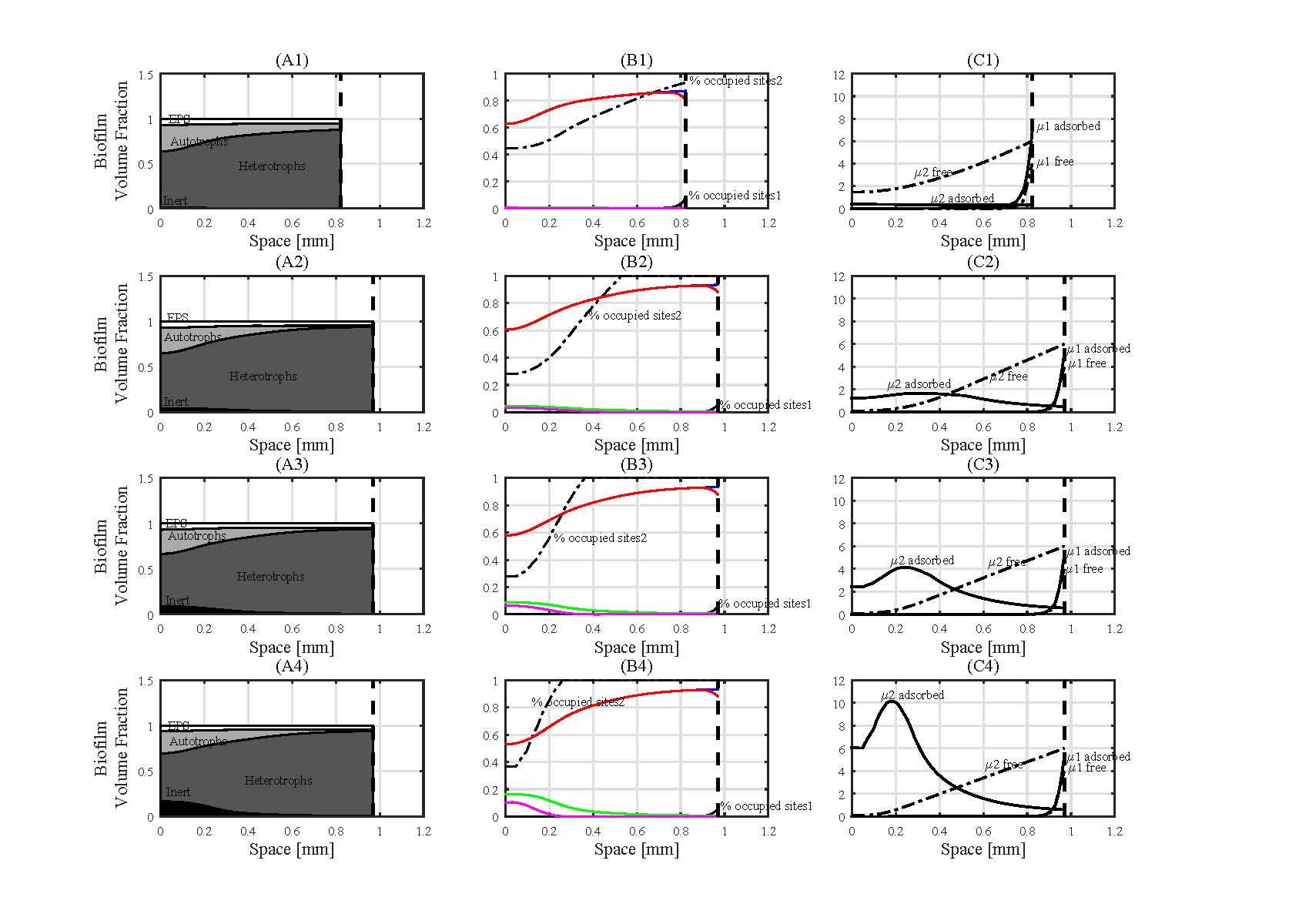}
 \caption{ Microbial species distribution (A1-A4); total (blue-line) binding sites fraction of Heterotrophic bacteria component, total (green-line) binding sites fraction of Inert component, free binding sites fractions (red-line) of Heterotrophic bacteria component, free binding sites fractions (magenta-line) of inert component, component  percentage of occupied sites of heterotrophic bacteria and inert components   (B1-B4); adsorbed and free metal profiles (C1-C4) after varied elapsed times: (A1,B1,C1) 1 days; (A2,B2,C2) 5 days; (A3,B3,C3) 10 days; (A4,B4,C4) 20 days. Free contaminant concentration is multiplied by a factor of $10^4$.}
          \label{f4.9}
 \end{figure*}

\begin{figure*}
 \includegraphics[width=1\textwidth]{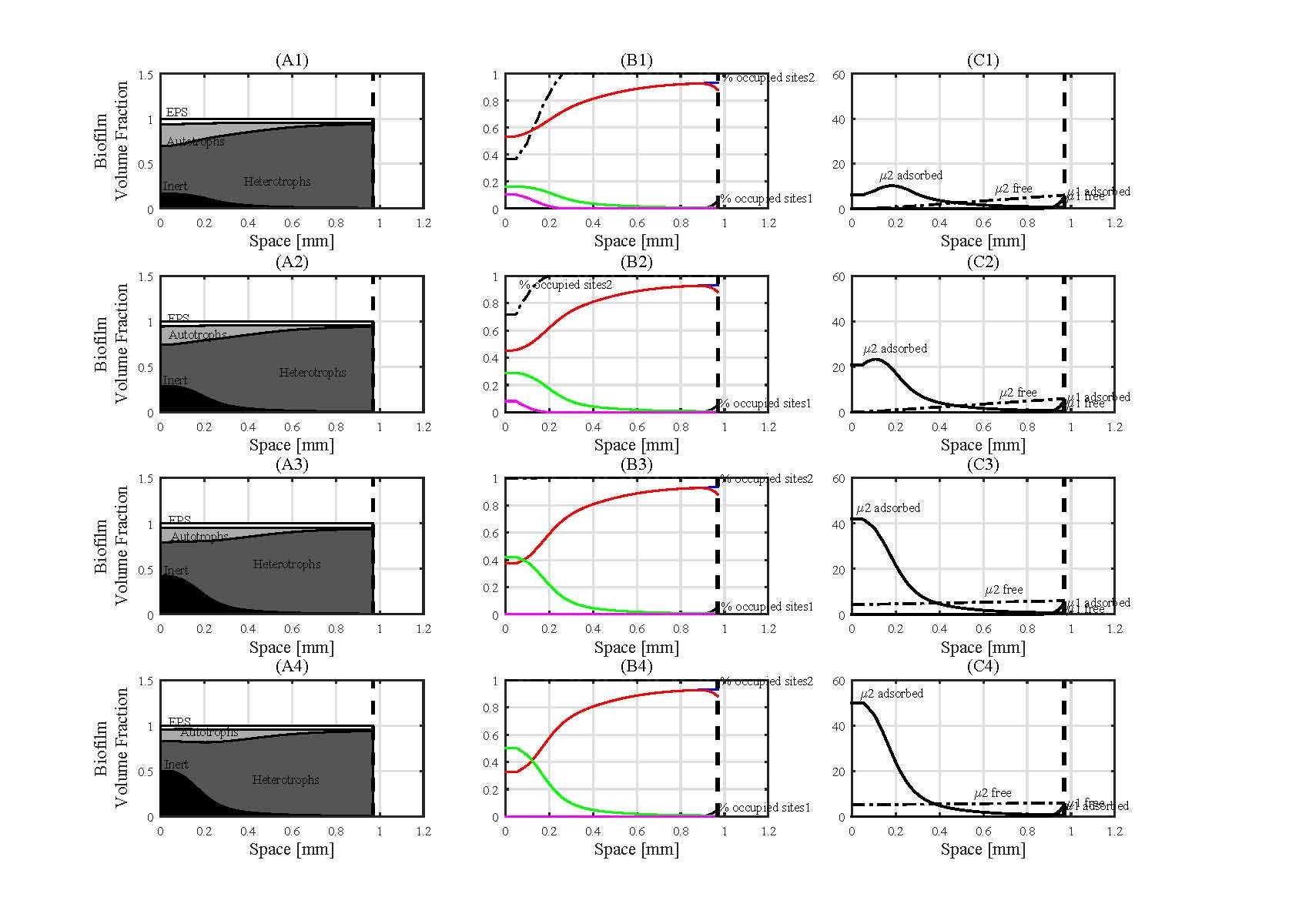}
 \caption{ Microbial species distribution (A1-A4); total (blue-line) binding sites fraction of Heterotrophic bacteria component, total (green-line) binding sites fraction of Inert component, free binding sites fractions (red-line) of Heterotrophic bacteria component, free binding sites fractions (magenta-line) of inert component, component  percentage of occupied sites of heterotrophic bacteria and inert components   (B1-B4); adsorbed and free metal profiles (C1-C4) after varied elapsed times: (A1,B1,C1) 30 days; (A2,B2,C2) 40 days; (A3,B3,C3) 70 days; (A4,B4,C4) 100 days. Free contaminant concentration is multiplied by a factor of $10^4$.}
          \label{f4.10}
\end{figure*}

This simulation scenario monitors over time the dynamics of the diffusion/reaction of the two contaminants in the heterotrophic-autotrophic biofilm system, figure \ref{f4.9} and \ref{f4.10}.
As in the previous simulation set, the heterotrophic bacteria predominate all over the biofilm with a maximum volume fraction on the free boundary. Interestingly, the inert fraction increases over time in the inner biofilm layers (figures \ref{f4.9} A1-A4 and \ref{f4.10} A1-A4).
The two free contaminant profiles $\mu_1$ and $\mu_2$ show immediately different trends (fig. \ref{f4.9}, C1) due to the different spatial distribution of the biofilm components on which they adsorb (fig. \ref{f4.9}, A1). Indeed, after one day simulation time, the inert concentration and thus the related free binding sites fraction  is very low (fig. \ref{f4.9}, B1 green line)  and can be rapidly saturated, as confirmed by the high percentage of saturated sites of inert, from almost $100\%$ in the external layer to about $40\%$ on the substratum (fig. \ref{f4.9}, B1). As expected the total amount of adsorbed contaminant $\mu_2$ is very low (fig. \ref{f4.9}, C1). On the contrary, the percentage of saturated sites on heterotrophic fraction differs from zero only in the external part of the biofilm due to the combined effect of higher concentration of binding sites and shear stress.

After 5 days, the percentage of  sites occupied by $\mu_2$ becomes $100\%$ in the external part of the biofilm while in the inner part decreases to $20\%$ (fig. \ref{f4.9}, B2) despite the concentration of adsorbed contaminant increases (fig. \ref{f4.9}, C2). This is due to the formation of new free binding sites of inert. More precisely, the rate of formation of new binding sites, due to the growth of inert fraction (fig. \ref{f4.9}, A2), is higher than the adsorption rate of contaminant $\mu_2$ on the same sites. Accordingly, the free metal profile of $\mu_2$ is linear in the external part of biofilm due to the absence of adsorption in this biofilm area (fig. \ref{f4.9}, C2).
Going on with time simulation (figures \ref{f4.9}, A3, B3, C3 and A4, B4, C4) the concentration of adsorbed contaminant $\mu_2$ on inert increases (figs \ref{f4.9}, C3 and C4), despite the percentage of occupied sites does not change substantially (figs \ref{f4.9}, B3 and B4). The invariance of the percentage of occupied sites can be ascribed to the quasi-equilibrium between the rate of the formation of sites and the adsorption rate.

After 40 days time simulation, despite the inert is not completely saturated, the percentage of occupied sites is close to the $100\%$ in the entire biofilm (fig \ref{f4.10}, B2). This means that the adsorption rate is higher than the formation rate of the new binding sites. Indeed as is possible to see in figures \ref{f4.9}, A2, A3 and A4, the biofilm has reached a quasi-stationary configuration in terms of biofilm component distribution and then the inert growth has decreased.

Figure (\ref{f4.10}, A3) shows the system configuration after $70$ days time simulation. The inert is completely saturated with the percentage of occupied sites equal to $100\%$ in the whole biofilm. The $\mu_2$ adsorbed profile reaches the final configuration (fig \ref{f4.10}, C3 and C4).

 \section{Conclusion}\label{n7}

 I have introduced a new mathematical model able to reproduce the dynamics of new emerging contaminants in multispecies biofilms. In particular, these contaminants can react with the biofilm matrix and be bio-sorbed by it or  they can be directly involved in the microbial methabolism. 
The model is based on a continuum description of the biofilm material and it is derived by using conservation principles.
The model is able to describe the growth of biofilm in terms of spatial distribution of microbial species and the others biofilm components, such as inert material and EPS. The diffusion of the substrates into the biofilm and their uptake from the microbial consortium is considered. The model takes into account the formation of binding sites for each biofilm component, as well as the occupation of the binding sites due to the contaminant sorption. The diffusion and the reaction of the free contaminants, their effect on the microbial growth and substrate degradation are directly taken into account.
The application of the model to the real special cases shows that the model can be used to assess dynamically the spatial distribution of one or more free and adsorbed contaminants on a specific biofilm component. In particular, the sorption process on a specific component has been considered. Simulation results show that it is strongly affected by the sorption properties of the specific biofilm component, such as binding sites density and adsorption constant  Moreover, the proposed model can be used to evaluate the dynamics of a contaminant that  can be sorbed on a microbial component of the biofilm and in the same time this contaminant can influence the biological activity of an other microbial group present in the biofilm.
Although the model applications are related to autotrophic-heterotrophic biofilms, the developed framework is general and could be applied to other biofilm systems as it captures many of the characteristics generally observed in biofilm sorption phenomena.

 \section{Acknowledgments}
The author would like to acknowledge Berardino D'Acunto and Maria Rosaria Mattei for the helpful conversations.


\vspace*{10cm}

\begin{thebibliography}{}

\bibitem{flemming2010biofilm} Flemming, H.C., Wingender, J. 2010. The biofilm matrix. {\em Nature Reviews Microbiology} \textbf{8}, 623:~633.

\bibitem{schleheck2009pseudomonas} Schleheck, D., Barraud, N., Klebensberger, J., Webb, J.S., McDougald, D., Rice, S.A., Kjelleberg, S. 2009.
Pseudomonas aeruginosa pao1 preferentially grows as aggregates in liquid batch cultures and disperses upon starvation. \textit{PloS one} \textbf{4}.

\bibitem{gadd2009biosorption} Gadd, G.M. 2009. Biosorption: critical review of scientific rationale, environmental importance and significance for pollution treatment. 
\textit{Journal of Chemical Technology and Biotechnology} \textbf{84}, 13:~28.

\bibitem{da2016copper} da Costa Waite, C.C., da Silva, G.O.A., Bitencourt, J.A.P., Sabadini-Santos, E., Crapez, M.A.C. 2016. 
Copper and lead removal from aqueous solutions by bacterial consortia acting as biosorbents. \textit{Marine pollution bulletin}.

\bibitem{lapworth2012emerging} Lapworth, D., Baran, N., Stuart, M., Ward, R. 2012 Emerging organic contaminants in
groundwater: a review of sources, fate and occurrence. \textit{Environmental pollution} \textbf{16}3, 287:~303.

\bibitem{d2015mathematical} D'Acunto, B., Esposito, G., Frunzo, L., Mattei, M., Pirozzi, F. 2015. Mathematical modeling
of heavy metal biosorption in multispecies biofilms. \textit{Journal of Environmental Engineering} 

\bibitem{van2003metal}van Hullebusch, E.D., Zandvoort, M.H., Lens, P.N. 2003. Metal immobilisation by biofilms: mechanisms and analytical tools. 
\textit{Reviews in Environmental Science and Biotechnology} \textbf{2}, 9:~33.

\bibitem{de2009failure} De Leenheer, P., Cogan, N. 2009. Failure of antibiotic treatment in microbial populations.
\textit{Journal of mathematical biology} \textbf{59}, 563:~579.

\bibitem{de2010senescence} De Leenheer, P., Dockery, J., Gedeon, T., Pilyugin, S.S. 2010. Senescence and antibiotic resistance in an age-structured population model.
\textit{Journal of mathematical biology} \textbf{61}, 475:~499.

\bibitem{rahman2015mixed} Rahman, K.A., Sudarsan, R., Eberl, H.J. 2015. A mixed-culture biofilm model with crossdiffusion.
\textit{Bulletin of Mathematical Biology} \textbf{77}, 2086:~2124.

\bibitem{AG-FCAA-2017}
A. Giusti,
On Infinite Order Differential Operators in Fractional Viscoelasticity, 
To appear in \textit{Fract. Calc. Appl. Anal.}, Vol. 20 (2017).
[E-print \href{https://arxiv.org/abs/1701.06350}{	arXiv:1701.06350} (2017)]

\bibitem{bartlett1996diffusion} Bartlett, P., Gardner, J. 1996. Diffusion and binding of molecules to sites within homogeneous thin films. 
\textit{Philosophical Transactions of the Royal Society of London A: Mathematical, Physical and Engineering Sciences} \textbf{354}, 35:~57.

\bibitem{art:rif.8} Wanner, O., Gujer, W. 1986. A multispecies biofilm model. \textit{Biotechnol Bioeng} \textbf{28} 314:~328.

\bibitem{art:rif.16} D'Acunto, B., Frunzo, L. 2011. Qualitative analysis and simulations of a free boundary problem for mulispecies biofilm models. 
\textit{Math Comput Model 53}, 1596:~1606.

\bibitem{art:rif.10} Dockery, J., Klapper, I. 2002. Finger formation in biofilm layers. \textit{SIAM J. Appl. Math.} \textbf{62}, 853:~869.

\bibitem{art:rif.17} D'Acunto, B., Esposito, G., Frunzo, L., Pirozzi, F. 2011. Dynamic modeling of sulfate reducing biofilms. \textit{Comput Math Appl} \textbf{62}, 2601:~2608.

\bibitem{art:rif.18} D'Acunto, B., Frunzo, L. 2012. Free boundary problem for an initial cell layer in multispecies biofilm formation. \textit{Appl Math Lett} \textbf{25}, 20:~26.

\bibitem{laspidou2002non} Laspidou, C.S., Rittmann, B.E. 2002. Non-steady state modeling of extracellular polymeric substances, soluble microbial products, and active and inert biomass. 
\textit{Water Research} \textbf{36}, 1983:~1992.

\bibitem{merkey2009modeling} Merkey, B.V., Rittmann, B.E., Chopp, D.L. 2009. Modeling how soluble microbial products (smp) support heterotrophic bacteria in autotroph-based biofilms. 
\textit{Journal of theoretical biology} \textbf{259}, 670:~683.

\end{thebibliography}
\end{document}